%% file: Sears_AOM_JMM2020.tex
\documentclass{tMAM2e}

\usepackage{epstopdf}
\usepackage[bookmarksnumbered,colorlinks,bookmarks,citecolor=blue,linkcolor=blue,urlcolor=blue,breaklinks,linktocpage]{hyperref}
\usepackage[all]{hypcap}
\usepackage{subfigure}
\usepackage{layouts}

\usepackage{booktabs} 
\usepackage[group-separator={,}]{siunitx} 
\usepackage[flushleft]{threeparttable} 
\usepackage{multirow} 
\usepackage{arydshln}
\usepackage{afterpage}

\theoremstyle{plain}

\theoremstyle{definition}

\theoremstyle{remark}

\title{Beneath (or beyond) the surface: Discovering voice-leading patterns with skip-grams}

\author{
	\name{David R. W. Sears\textsuperscript{a}$^{\ast}$\thanks{$^\ast$Corresponding author. Email: david.sears@ttu.edu},
		and Gerhard Widmer\textsuperscript{b}
	}
	\affil{\textsuperscript{a}J. T. \& Margaret Talkington College of Visual \& Performing Arts, Texas Tech University, Lubbock, Texas}
	\affil{\textsuperscript{b}Institute of Computational Perception, Johannes Kepler University, Linz, Austria}
	\received{}
}

\begin{document}

\maketitle

\begin{abstract}
Recurrent voice-leading patterns like the Mi-Re-Do compound cadence (MRDCC) rarely appear on the musical surface in complex polyphonic textures, so finding these patterns using computational methods remains a tremendous challenge. The present study extends the canonical $n$-gram approach by using \textit{skip-grams}, which include sub-sequences in an $n$-gram list if their constituent members occur within a certain number of skips. We compiled four data sets of Western tonal music consisting of symbolic encodings of the notated score and a recorded performance, created a model pipeline for defining, counting, filtering, and ranking skip-grams, and ranked the position of the MRDCC in every possible model configuration. We found that the MRDCC receives a higher rank in the list when the pipeline employs 5 skips, filters the list by excluding $n$-gram types that do not reflect a genuine harmonic change between adjacent members, and ranks the remaining types using a statistical association measure.
\end{abstract}


\section{Introduction}

Pattern discovery is an essential task in many fields, but particularly so in that branch of criticism concerned with the theory and analysis of music. According to \citet[83]{Simon1993}, ``one of the purposes of analyzing musical structure and form is to discover the patterns that are explicit or implicit in musical works.'' \citet[19]{Herskovits1941} would seem to agree, arguing that ``the peculiar value of studying music ... is that, even more than other aspects of culture, its patterns tend to lodge on the unconscious level." \citet{Margulis2013,Margulis2014} and \citet{Fitch2006} have even suggested that the predilection for pattern repetition distinguishes music from language more than any other design feature. 

From this vantage point, it should be no surprise that the tonal cadence continues to receive so much attention in contemporary scholarship. As a highly replicated closing pattern appearing at the ends of phrases, themes, and larger sections, the cadence provides perhaps the clearest instance of phrase-level schematic organization in the tonal system \citep{Sears2018}. \citet[43]{Dunsby1980} has argued, for example, that cadences remain ``one of the few consistently patterned aspects of musical structure," while \citet[105]{Sanguinetti2012} contends that cadences represent ``the first, most elementary of tonal structures," providing a flexible scaffold on which to build increasingly complex diminutions spanning phrases, sections, and entire pieces.

\begin{figure}[t!]
	\centering{}
	\fontsize{8pt}{10pt}\selectfont
	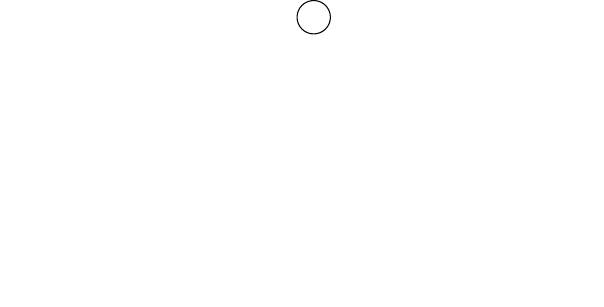
	\caption{Beethoven, Op. 26, i, mm. 15-16. Melodic scale-degrees and Roman numeral annotations appear above and below, respectively.}
	\label{fig:beethoven_simple}
\end{figure}

According to \citet{Meyer2000b}, the Mi-Re-Do compound cadence (MRDCC) is perhaps the most important and highly replicated closing pattern in Western music of the common-practice period (1610-1900). The passage in Figure \ref{fig:beethoven_simple}, which closes the main theme in the first movement of Beethoven's Op. 26, presents the underlying voice-leading scaffold. The MRDCC is a three-stage formula that resolves a six-four embellishment of dominant harmony to root position before proceeding to tonic harmony. Contemporaneous scholars from the Neapolitan tradition referred to such patterns as \textit{cadenze composte} (or \textit{compound cadences}) because they allot two metrical units to the dominant (e.g., V$^6_4$--V$^5_3$--I) \citep{Sanguinetti2012},\footnote{By comparison, \textit{cadenze semplici} (or \textit{simple cadences}) receive one metrical unit (e.g., ii$^6$--V--I), and \textit{cadenze doppie} (or \textit{double cadences}) receive four units (e.g., V$^5_3$--$^6_4$--$^5_4$--$^5_3$--I).} but the term \textit{cadential six-four} is now commonplace in contemporary pedagogical texts \citep{Aldwell2003, Clendenning2016, Kostka2018}. Although the compound cadence may support a number of contrapuntal patterns in the melody (e.g., Do-Ti-Do, Sol-Fa-Mi, etc.), \citet[235]{Meyer2000b} suggested that the Mi-Re-Do stepwise melodic descent reflects ``a profound stylistic change'' in the history of Western music, from the contrapuntal principles of Renaissance music, to the syntactic principles associated with music of the Baroque and Classical periods. 

Thus, the MRDCC is assumed by many to be the quintessential tonal closing schema for music of the common-practice period, ``a microcosm which summarizes the essential features ... of the work it closes" \citep[iii]{Casella1924}. And yet, despite the remarkable ubiquity and diversity of patterns like the MRDCC in both Western and non-Western tonal musics \citep{Meyer2000b}, data-driven methods for the discovery, classification, and prediction of recurrent temporal patterns in polyphonic corpora have yet to gain sufficient traction in music research. This fact owes in large part to presumed limitations associated with string-based methods, which typically divide a musical corpus into \textit{contiguous} sub-sequences of $n$ events (called $n$-grams), and so mistakenly assume that note or chord events on the musical surface depend only on their immediate neighbors. To be sure, much of the world's music is hierarchically organized such that certain events are more stable or important than others, and so non-contiguous events often serve as focal points in the sequence \citep{Gjerdingen2014}. As a consequence, existing string-based methods often fail to identify patterns featuring non-contiguous events, a limitation \citet{Collins2014} have called the \textit{interpolation problem}.

\begin{figure}[t!]
	\centering{}
	\fontsize{8pt}{10pt}\selectfont
	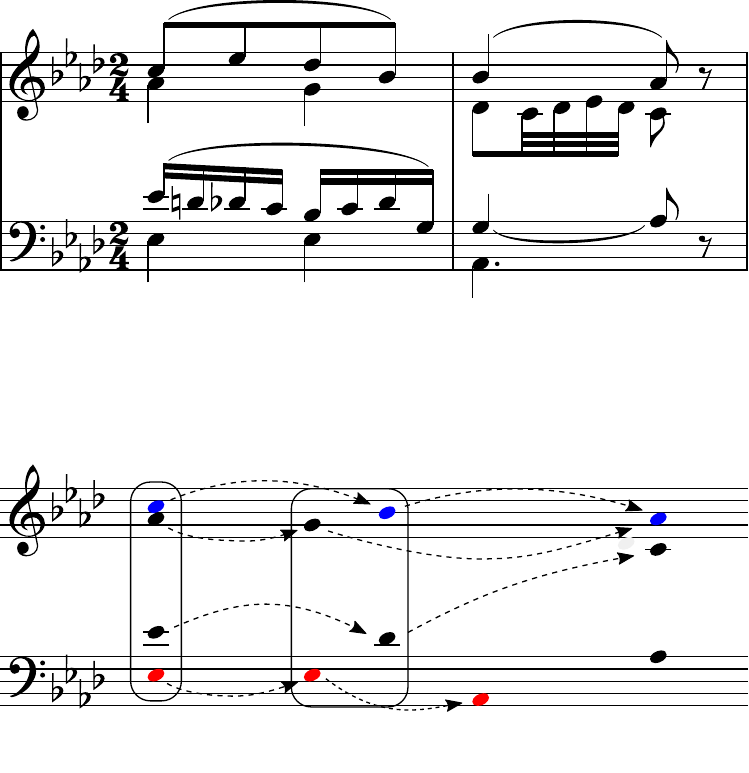
	\caption{(a) Beethoven, Op. 10, No. 1, ii, mm. 15-16. (b) A relational network for the cadential schema, with melodic scale-degrees and Roman numeral annotations appearing above and below, respectively.}
	\label{fig:beethoven_complex}
\end{figure}

By way of example, consider the MRDCC from the closing measures of the main theme in the second movement of Beethoven's Op. 10, No. 1, shown in Figure \ref{fig:beethoven_complex}a. The passage is in many respects a conventional exemplar of the MRDCC, but the underlying voice-leading scaffold is obscured by embellishing tones that promote smooth voice-leading within each voice and exchange the `core' tones of the MRDCC between voices. The network of relations depicted in Figure \ref{fig:beethoven_complex}b reduces the passage to the core tones of the MRDCC, with the bass and soprano melodies shown in red and blue, respectively. The connections between note events forming harmonic complexes (i.e., chords) appear inside boxes, the connections between note events within each contrapuntal voice receive dotted and curved arrows, and the connections between the harmonic complexes themselves receive double-lined arrows. Whether listeners would reduce this complex passage to the scaffold in Figure \ref{fig:beethoven_complex}b is itself an open question, but presumably the core tones of the MRDCC (co-)occur with sufficient frequency to justify the melodic scale-degree and Roman numeral annotations that appear above and below the network, respectively. And yet, contiguous string-based methods would fail to recover this structure.

To uncover potentially remote relationships between words in natural language corpora, researchers in corpus linguistics and natural language processing (NLP) have developed \textit{skip-grams} \citep{Guthrie2006}, an alternative string-based method that includes sub-sequences in an $n$-gram distribution if their constituent members occur within a certain number of skips. To identify ``characteristic and frequently recurrent word combinations" that stand in a more flexible relationship to one another, such as \textit{knock ... door} (e.g., \textit{knock at the door}, \textit{knock on the door}, etc.), for example, collocation discovery algorithms often rely on the following analysis pipeline \citep{Evert2008}: 
\begin{itemize}
	\item identify recurrent word combinations using skip-grams (\textit{skip})
	\item count instances of each resulting skip-gram type (\textit{count})
	\item filter out irrelevant types (\textit{filter})
	\item rank the remaining types (\textit{rank})
\end{itemize}

Determining the appropriate configuration of methods is a difficult empirical problem, however. The researcher must select the length of the skip boundary, the precise method of counting, the criteria for filtering, and finally, the statistical ranking measure. Depending on the number of methods selected for each stage of the pipeline, the resulting algorithm can produce thousands of model configurations. Thus, previous studies have evaluated collocation discovery algorithms by identifying the configuration of methods that optimizes the ranks of collocations that were previously identified by expert annotators \citep{Petrovic2010}. Simply put, model configurations that produce higher ranks for a given set of collocations are assumed to be more suitable to the task.  

The goal of the present study is to adapt this analysis pipeline for the discovery of recurrent voice-leading patterns like the MRDCC in complex polyphonic textures. To that end, we have selected a number of methods for each stage of the pipeline and then identified the configuration of methods that optimizes the rank of the MRDCC in a corpus of polyphonic music. If the MRDCC is indeed one of the most highly replicated and characteristic patterns in music of the common-practice period \citep{Meyer2000b}, optimizing its rank will allow us to determine the best-performing model configuration and identify other relevant voice-leading types. The corpus examined here consists of four data sets of Western classical music and includes both string quartet and piano textures. What is more, since previous musical pattern discovery algorithms for polyphonic corpora have relied on both score-based temporal features measured in metrical time (e.g., beats) and audio-based features measured in clock time (e.g., seconds), each data set features symbolic representations of both the notated score and a recorded performance.

We begin in Section \ref{sec:repschemes} by briefly reviewing pattern discovery methods in music research and then describing the voice-leading type (VLT) representation scheme, an optimally reduced chord typology that models all possible combinations of note events in a polyphonic data set, but that reduces the number of distinct chord types based on music-theoretic principles. Next, Sections \ref{sec:defineskipgrams}-\ref{sec:rankskipgrams} present methods for defining, counting, filtering, and ranking skip-grams. Section \ref{sec:methods} describes the corpora and evaluation procedure used in the present research, and Section \ref{sec:results} presents the results of the model evaluation and examines the top-ten voice-leading patterns from the optimal model configuration. Finally, we conclude in Section \ref{sec:conclusion} by considering limitations and directions for future research.

\section{Representation schemes}\label{sec:repschemes}

Corpus studies in music research often privilege the \textit{note} event, examining features like chromatic pitch \citep{Pearce2004}, melodic interval \citep{Vos1989}, or chromatic scale degree \citep{Margulis2008}. Identifying \textit{composite} events like triads and seventh chords in polyphonic textures is considerably more complex, as the number of distinct $n$-note combinations is often enormous. To resolve this issue, previous corpus studies have either reduced the surface to a sequence of harmonies from a specific chord typology and used string-based methods to identify relevant sub-sequences, or abandoned string-based methods in favor of point-set (or \textit{geometric}) methods. In the first (string-based) approach, researchers select a chord typology \textit{a priori} (e.g., the Roman numeral system, figured bass nomenclature, or pop chord symbol notation), and then identify chord events using either human annotators \citep{Declercq:2011,Tymoczko:2011,Burgoyne2012}, or rule-based computational classifiers trained on homorhythmic genres, where conventional chord progressions are more likely to occur on the surface (e.g., Bach chorales) \citep{Temperley:1999,Rowe:2001,Cambouropoulos2016}. Yet unfortunately, existing typologies depend on a host of assumptions about the sorts of simultaneous relations the researcher should privilege (e.g., triads and seventh chords), and, depending on the corpus, may also require additional information about the underlying tonal context, which again must be inferred either during transcription \citep{Margulis2008}, or using some automatic (key-finding) method \citep{White2015}. In the second (geometric) approach, researchers have resolved the interpolation problem by representing note events as points in a multidimensional space \citep{Meredith2002,Collins2016}. However, the geometric approach generally does not extend to prediction tasks, where string-based methods excel (e.g., $n$-gram models). 

To identify chord progressions in polyphonic corpora using string-based methods, previous studies have constructed composite chord events from simpler combinations of simultaneous note events. To that end, many software frameworks perform a \textit{full expansion} of the symbolic encoding, which duplicates overlapping note events at every unique onset time \citep{Sears2017}. Thus, each unique onset time is represented as a vertical slice consisting of all sounding note events. The Humdrum software framework calls this technique \textit{ditto} \citep{Huron1993}, while Music21 calls it \textit{chordifying} (or \textit{salami-slicing}) \citep{Cuthbert2010}.\footnote{See \citet{Sears2017} for a worked example.} Although expansion fails to identify composite chord events featuring non-overlapping members (e.g., a chord that appears in an Alberti bass pattern), it is still less likely to under-partition more complex polyphony compared to other partitioning methods \citep{Conklin2002}, so we adopt this technique here. 

Chord onsets from the expanded encoding are typically represented according to the simultaneous relations between their note-event members (e.g., vertical intervals) \citep{Sears2016}, the sequential relations between their chord-event neighbors (e.g., melodic intervals) \citep{Conklin2002}, or some combination of the two \citep{Quinn2010}. The skip-gram method can model any of these schemes, but we have adopted the \textit{voice-leading type} (VLT) representation developed by \citet{Quinn2010} and \citet{Quinn2011}, which produces an optimally reduced chord typology that still models every possible combination of note events. For our purposes, the VLT scheme consists of an ordered tuple ($S, T, I$) for each chord onset in the expanded encoding, where $S$ is a set of up to three intervals above the bass, $T$ is the interval between the bass and highest instrumental part, and $I$ is the melodic interval from the preceding bass note to the present one. All intervals are measured in semitones modulo the octave. Thus, the value of each interval class is either undefined (denoted by $\perp$), or represents one of twelve possible interval classes, where 0 denotes a perfect unison or octave, 7 denotes a perfect fifth, and so on. The inclusion of the simultaneous relation(s) in $S$ and $T$ therefore ensures that the most common VLTs will have analogues in conventional chord typologies (e.g., triads and seventh chords), while the inclusion of the sequential relation in $I$ ensures that the resulting VLT sequences remain invariant to their underlying tonal context, yet still retain enough voice-leading information to reveal how they progress over time. 

Because the VLT representation makes no distinction between chord tones and non-chord tones, the syntactic domain of voice-leading types is very large. Typically, the precise location and repeated appearance of a given pitch class is assumed to be irrelevant to the identity of a given sonority \citep{Quinn:2010}, so we have excluded pitch class repetitions (i.e., voice doublings) and allowed permutations in $S$. By allowing permutations, the major triads $\langle4, 7, 0\rangle$ and $\langle7, 4, 0\rangle$ reduce to $\langle4, 7, \perp\rangle$. Similarly, by eliminating repetitions, the major-minor seventh chords $\langle4, 4, 10\rangle$ and $\langle4, 10, 10\rangle$ reduce to $\langle4, 10, \perp\rangle$.

Following the notation scheme $\langle S* \rangle$[$I$], the MRDCC in Figure \ref{fig:beethoven_simple} would receive the following encoding:

\begin{equation*}
\langle5, 9*, \perp\rangle \enspace [0] \enspace \langle4, 7*, 10\rangle \enspace [5] \enspace \langle4, \perp, \perp\rangle 
\end{equation*}

An asterisk denotes the interval class of the highest voice, $T$. VLT members without an asterisk in $S$ indicate that the highest voice doubles the bass at the unison or octave. Thus, the MRDCC is a 3-gram consisting of three chords, $S_{\text{1}}$, $S_{\text{2}}$, and $S_{\text{3}}$, with two melodic interval classes, $I_{\text{1}}$ and $I_{\text{2}}$, connecting those chords. 

One limitation of the VLT scheme is that several patterns could reflect the same underlying type. A given exemplar of the MRDCC, for example, could omit the seventh in the penultimate dominant, include the fifth in the final tonic, or omit the third and fifth in the final tonic. All of these variants would receive a distinct $n$-gram type in the VLT scheme. To simplify the evaluation procedure, we elected to evaluate the analysis pipeline using the variant of the MRDCC found in Figures \ref{fig:beethoven_simple} and \ref{fig:beethoven_complex} and described by \citet{Meyer2000b}, which includes a complete dominant seventh chord that resolves to an incomplete tonic that omits the fifth.

\section{Defining skip-grams}\label{sec:defineskipgrams}


Researchers typically discover recurrent patterns by dividing the corpus into contiguous sub-sequences of cardinality $n$ (called $n$-grams), and then counting the number of instances (or \textit{tokens}) associated with each distinct $n$-gram \textit{type} in the corpus. When $n$ is small, tokens from the list of $n$-gram types typically receive prefixes (e.g., unigrams, bigrams, trigrams, etc.), but longer $n$-grams are represented by the value of $n$ (e.g., 5-grams).

\subsection{Contiguous {\it n}-grams}

Identifying contiguous $n$-grams is relatively straightforward. If each composition $m$ consists of a contiguous sequence of VLTs, let $k$ represent the length of the sequence, and let $C$ denote the total number of compositions in the corpus. The number of contiguous \textit{n}-gram tokens in the corpus is 
\begin{equation}\label{eq:1}
\displaystyle\sum_{m=1}^{C} k_m-n+1
\end{equation}
This formula implies that the total number of tokens is necessarily smaller than the total number of events in the sequence when $n>1$.

\subsection{Non-contiguous (skip) {\it n}-grams}

\begin{figure}
	\centering
	\def\svgwidth{.7\columnwidth}
	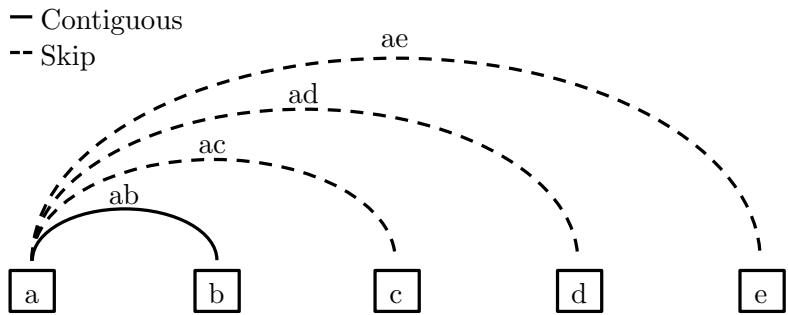
	\caption{A 5-event sequence, with arcs denoting all contiguous (solid) and non-contiguous (dashed) 2-gram tokens featuring $a$ as a member.}
	\label{fig:non_contiguous}
\end{figure}

String-based methods using contiguous $n$-grams only consider directly adjacent events. Without this restriction, the number of associations between events in the sequence necessarily explodes in combinatorial complexity as $n$ and $k$ increase. Figure \ref{fig:non_contiguous} depicts the 2-gram tokens for a 5-event sequence involving $a$ as a member, with solid and dashed arcs denoting contiguous and non-contiguous relations, respectively. The number of tokens that include all possible contiguous and non-contiguous relations is given by the combination equation. Thus, the number of tokens can very quickly become unwieldy as $n$ and $k$ increase: a 20-event sequence contains 15,504 5-grams, for example. 



\subsubsection{Fixed skip {\it n}-grams}\label{sec:fixedskip}

To overcome the computational processing and storage limitations associated with counting tokens in this way, \citet{Guthrie2006} limited the investigation to what \citet{Sears2017} have called \textit{fixed-skip} \textit{n}-grams, which only include \textit{n}-gram tokens if their constituent members occur within a fixed number of skips $t$. Shown in Figure \ref{fig:non_contiguous}, $ac$ constitutes a 1-skip token (i.e., $t=1$), while $ad$ and $ae$ constitute 2- and 3-skip tokens, respectively. Using the skip-gram method, \citet{Sears2017} showed that the inclusion of skip-grams reduces sparsity in higher-order \textit{n}-gram count distributions, thereby improving the accuracy of probabilistic models tasked with melodic and harmonic prediction, classification, and pattern discovery.  

\subsubsection{Variable skip {\it n}-grams}\label{sec:variableskip}

In music corpora, temporal characteristics like rhythmic duration and metric position play an essential role in the realization and reception of musical works. \citet{Fraisse1982} found, for example, that the upper boundary under which listeners can group successive events into temporal sequences is around 2s. Thus, as an alternative to the fixed-skip method, we also include \textit{variable-skip} \textit{n}-grams, which include \textit{n}-gram tokens if the inter-onset interval(s) (IOI) between their constituent members occur within a specified time interval (e.g., 2s), regardless of the number of intervening events between any two of them.

\section{Counting skip-grams}\label{sec:countskipgrams}

Count distributions of $n$-gram types assign equal weight to each encountered token regardless of the perceived salience or memorability of its members. In this sense, count is simply a binary (indicator) function denoting the presence (or absence) of a given $n$-gram token, producing the values $\{0,1\}$. Thus, an $n$-gram token featuring long, irregular inter-onset intervals (IOIs) between adjacent members would receive the same count \textit{value} as one consisting of relatively short, periodic IOIs (i.e., each would be counted once). Key-finding studies have generally reported improved performance when using weighting functions, however, which produce count values on the real unit interval $[0, 1]$, thereby adjusting the final count to ensure that tokens featuring more salient or memorable members will receive values closer to 1. For example, \citet{Huron1993} have shown that key-finding improves when the temporal sequence is weighted by an exponential decay function that simulates the effects of sensory (or echoic) memory during music listening. 

To weight $n$-gram tokens in this way, previous studies have typically extracted score-based temporal features measured in metrical time (e.g., beats), or audio-based features measured in clock time (e.g., seconds). The Krumhansl-Schmuckler algorithm, for example, weights the counts associated with a given pitch-class distribution by the rhythmic duration of the encountered events, thereby rewarding events with longer durations in the final count \citep{Krumhansl1990}. Listeners rarely encounter metronomic performances of the score in everyday listening, however. Ideally, pattern discovery algorithms would take as input audio recordings of musical performances, but given the difficulties associated with automatic transcription \citep{Benetos2013}, we have elected to extract pitch-based features from the symbolic encoding, but use temporal features encoded from real performances of the corpus.\footnote{This decision requires actual recordings that are precisely aligned to their respective symbolic scores. The musical corpora to be used in this study (see Section \ref{sec:corpora}) are of this kind.} This dual encoding scheme allows us to examine weighting functions from the music perception and cognition literature in the pattern discovery pipeline without the loss in performance associated with automatic transcription. Thus, in addition to a simple binary count function, the following four weighting functions produce count values on the interval $[0,1]$ for each $n$-gram token, $\tau$: periodicity, resonance, proximity, and resonant periodicity. 


\subsection{Periodicity}

Listeners are predisposed to finding a regular (i.e., isochronous) or periodic pulse in auditory stimuli \citep{McAdams2002}, so the goal of a periodicity function is to privilege tokens whose members feature periodic IOIs.  \citet{Large1999} developed a computational model of meter perception that synchronizes a bank of oscillators to the periodicities in an external signal. They describe this process using a sine circle map \citep{Glass1988}, which entrains to a periodic signal using a discrete-time formalism. Each cycle of the sine map determines the period of oscillation, $p$. The phase, $\phi$, of each $n$-gram event member, $\text{onset}_i$, is the position of the oscillation around the cycle. By defining the onset time at which an event is expected to occur as $\text{onset}_x$, and defining $\phi(\text{onset}_x)$ as 0, \citet{Large1999} produce the following relation:
\begin{equation*}\label{eq:periodicity1}
\phi(\text{onset}_i) = \frac{\text{onset}_i-\text{onset}_x}{p} 
\end{equation*}

It is also possible to determine the phase of the next onset, $\phi(\text{onset}_{i+1})$, from the phase of the current onset, $\phi(\text{onset}_i)$, the IOI between the event members, $\text{IOI}_s$, where $\text{IOI}_s=\text{onset}_{i+1}-\text{onset}_{i}$, and the period, $p$:   
\begin{equation}\label{eq:periodicity2}
\phi_{i+1} = \phi_{i}+\frac{\text{IOI}_s}{p} \quad (\text{mod}_{-0.5,0.5} \, 1) 
\end{equation}
The expression, $(\text{mod}_{-0.5,0.5} \, 1)$, maps the event members in each token onto the circle by taking the remainder after division by 1 and then remapping the interval $(0.5,1)$ to the interval $(-0.5,0)$ \citep[p. 126]{Large1999}. This equation produces a circle map, with the period of oscillation acting as an autonomous referent, or clock. The phase of the clock describes the onsets of the event members in token $\tau$ with respect to the candidate period, $p$, with each IOI$_s$ in $\tau$ serving as a candidate period in the final periodicity weighting function.

To calculate the periodicity of $\tau$ on the interval $[0,1]$, we estimate the mean vector length coefficient for each candidate period from the relative phases estimated in Equation \eqref{eq:periodicity2}:
\begin{equation}\label{eq:periodicity3}
W_{\text{period}}(\tau_p) = \frac{1}{n}\displaystyle\sum_{i=1}^{n}\cos2\pi(\phi_{i}-\overline{\phi})
\end{equation}
The mean vector length is a circular statistic that measures dispersion about the mean, but it is also used as a measure of synchronization strength \citep{Goldberg1969}. 

To calculate the periodicity weight for $\tau$, we estimate $W_{\text{period}}$ for each candidate period, $\tau_p$, and then take the minimum of these estimates. Thus, $W_{\text{period}}$ ensures that tokens featuring regular IOIs between event members will receive a value of 1, whereas irregular or aperiodic tokens will receive values closer to 0.

\subsection{Resonance}

Listeners tend to perceive a regular pulse at a preferred tempo range from 80 to 160 beats-per-minute (BPM) \citep{Moelants2002}. To model the influence of these resonant periodicities for $\tau$, we use the parameterized resonance model developed by \citet{vanNoorden1999}. They estimate the effective resonance amplitude, which we will call $W_{\text{res}}$, using a damped harmonic oscillator:
\begin{equation}\label{eq:resonance1}
W_{\text{res}}(\tau) = \frac{1}{\sqrt{(f^2_0 - p^2)^2 + \beta p^2}} - \frac{1}{\sqrt{f^4_0+p^4}}
\end{equation}
Here, $f_0$ is the resonant period (2 Hz, or 0.5s), $\beta$ is the damping constant (1.12), and $p$ is the candidate period determined by $W_{\text{period}}$. Each of the first two parameters was modeled in \citet{vanNoorden1999} to account for the tapping data in \citet{Handel1981}, in which participants were asked to tap regularly (i.e., isochronously) to polyrhythmic sequences (e.g., 2:3, 3:4, etc.) presented at various tempi. Thus, $n$-gram tokens whose IOIs correspond closely with the resonant periodicity of 0.5s receive higher weights than those with IOIs much shorter or longer than 0.5s.

\subsection{Proximity}

In a key-finding study, \citet{Huron1993} modeled memory decay for chord sequences using an inverse exponential function with a half-life of 1s to account for the decay in sensory memory resulting from the duration of the IOI between two chords. 
\begin{equation}\label{eq:proximity}
W_{\text{prox}}(\tau) = \displaystyle \frac{1}{n-1} \displaystyle\sum_{i=2}^{n}2^{(\text{onset}_{i-1} - \text{onset}_{i})} 
\end{equation}

The proximity weighting function, $W_{\text{prox}}$, represents the average decay in the IOIs between adjacent members. In this case, tokens featuring synchronous members receive a value of 1 (maximal contiguity), whereas tokens whose IOIs approach 2s receive values closer to 0 (minimal contiguity).\footnote{The optimum half-life for sensory memory varies from anywhere between 0.5s and 3s in the experimental literature, but we have elected to retain the half-life of 1s obtained in their study \citep[pp. 165-166]{Huron1993}.}

\subsection{Resonant Periodicity}

\citet{Parncutt1994} modeled the overall salience of a pulse sensation as the product of the periodicity (which in his model was called \textit{pulse-match salience}), and the resonance (or \textit{pulse-period salience}) of a given sequence. Thus, we also weight each $n$-gram token according to a resonant periodicity function using the same equation:
\begin{equation}\label{eq:resonantperiodicity}
W_{\text{res\_period}}(\tau) =W_{\text{period}}(\tau) \cdot W_{\text{res}}(\tau)
\end{equation}
Here, $W_{\text{res\_period}}$ produces higher weighted counts for $n$-gram tokens whose members feature regular (or periodic) IOIs of approximately 0.5s.

\section{Filtering skip-grams}\label{sec:filterskipgrams}

According to \citet{Manning1999}, the most important step in any collocation discovery algorithm is filtering, the goal of which is to remove irrelevant $n$-gram types from the final list. Corpus linguists typically exclude types either because they contain too few tokens to justify closer examination, or because they reflect parts of speech (POS) or syntactic categories ``that are rarely associated with interesting linguistic expressions" \citep[31]{Manning1999}. Applying a frequency filter is simply a matter of excluding types whose counts do not meet the specified count threshold, but POS filters require domain-specific knowledge about the syntactic categories that characterize the language(s) in a given text corpus. To that end, researchers often employ automatic annotation algorithms for very large corpora where manual tagging is no longer feasible.

To extend these methods to a corpus of voice-leading patterns, we will apply a frequency filter, a harmony filter analogous to the POS filters used in corpus linguistics, and both filters in combination.

\subsection{Frequency}

Selecting a threshold for frequency filters is unfortunately somewhat arbitrary. Previous studies involving text copora typically suggest a minimal threshold of $f\geq3$ or $f\geq5$, with higher thresholds often leading to even better results in practice \citep{Evert2008}. Given the size of the fixed and variable skip windows selected for this study (up to 8 skips and 2s, respectively), we elected to set a threshold of $f\geq10$ tokens for each $n$-gram type. 

\subsection{Harmony}

The musical surface contains considerable surface repetition, particularly for genres featuring complex polyphonic textures (e.g., string quartets, piano sonatas, symphonies, etc.). In a corpus study examining two-chord tonal progressions in Haydn's string quartets, for example, \citet{Sears2019} found that 9 of the top 10 bigram types were exact repetitions of the same chord type (e.g., I--I). Perhaps worse, a significant portion of the chord tokens on the expanded surface consisted of fewer than three distinct pitch events. By comparison, voice-leading progressions containing more than one distinct harmony were much less common. A harmony filter might therefore exclude $n$-gram types if they do not include a genuine harmonic (i.e., pitch) change between primarily tertian sonorities \citep[231]{Meyer2000b}. Thus, we have excluded $n$-gram types if (1) any chord member contains only one distinct pitch class (\textit{monophony}); (2) no chord member contains at least three distinct pitch classes (\textit{polyphony}); (3) all chord members feature the same pitch class in the bass (\textit{change of harmony}); (4) adjacent chord members share the same pitch class in the bass and any interval classes above the bass (\textit{similarity}). The first two criteria privilege tertian sonorities, while the latter two emphasize pitch change between adjacent chord members.  

%
%

\subsection{Both}

The `both' filtering method applies the frequency and harmony filters simultaneously. 
%
%

\section{Ranking skip-grams}\label{sec:rankskipgrams}

Ranking $n$-grams by their count assumes that the most common patterns will be the most important. But as \citet[56]{Temperley2018} points out, ``it is sometimes unclear whether such patterns are true `schemata' in the minds of ... musicians (and listeners) ... After all, a [Rock] progression such as I--IV--V--IV consists entirely of common chords and common harmonic moves; it would be surprising if it did \textit{not} occur." Simply put, frequent events are more likely to co-occur just by chance \citep[5]{Evert2008}. 

To address this issue, corpus linguists have developed \textit{association} (or \textit{attraction}) \textit{measures} (AMs) that rank $n$-gram types according to the statistical associations between their constituent members \citep{Evert2008}. The logic behind these measures is that a linguistic expression whose \textit{observed} frequency within the corpus is greater than chance---as measured by, for example, an \textit{expected} frequency associated with the joint probability of their constituent members---should receive a higher rank in the final list.\footnote{Observed and expected frequencies in AMs can refer to probability estimates. In such cases, they are often called \textit{relative} (as opposed to \textit{absolute}) frequencies. Thus, if the observed absolute frequency for a given type is 15 and the total number of tokens in the data set is 100, the relative frequency for that type is .15.} Such expressions are therefore deemed to be more salient, important, or memorable because their individual words statistically point to (or signify) the expression in toto. A bigram like \textit{weapons of} undoubtedly cues readers to expect the consequent bigram \textit{mass destruction}, for example, suggesting statistically attracted events form large-scale, idiomatic expressions. Thus, AMs rank each $n$-gram type not by its count, but by some method of probabilistic inference. 

AMs typically base their scores on contingency tables that represent the cross-classification between $n$ events \citep{Evert2008}. Shown in Figure \ref{fig:contingency}, the $2\times2$ contingency table on the left represents the bigram tokens containing both chord$_1$ and chord$_2$ ($O_{11}$), chord$_1$ but not chord$_2$ ($O_{12}$), chord$_2$ but not chord$_1$ ($O_{21}$), and neither of the two chords ($O_{22}$). The sum of these observed frequencies is equal to the total number of tokens in the corpus, $N$. The marginal frequencies denote the row and column sums. $R_1$, for example, corresponds to the number of tokens containing chord$_1$.  

\begin{figure}
	\def\arraystretch{3.5}
	\begin{minipage}[t]{.5\linewidth}
		\centering
		\begin{tabular}{r c c l}
			& \multicolumn{1}{|c|}{$=$ chord$_2$}     & \multicolumn{1}{c|}{$\neq$ chord$_2$} &            \\
			\cline{1-3}
			\multicolumn{1}{r}{$=$ chord$_1$}    & \multicolumn{1}{|c|}{$O_{11}$}   & \multicolumn{1}{c|}{$O_{12}$} & $= R_1$        \\
			\cline{1-3}
			\multicolumn{1}{r}{$\neq$ chord$_1$}   & \multicolumn{1}{|c|}{$O_{21}$}  & \multicolumn{1}{c|}{$O_{22}$}  & $= R_2$      \\
			\cline{1-3}
			& $= C_1$ & $= C_2$ & $= N$ \\
		\end{tabular}%
	\end{minipage}
	\begin{minipage}[t]{.5\linewidth}
		\centering
		\begin{tabular}{r c c }
			& \multicolumn{1}{|c|}{$=$ chord$_2$}     & \multicolumn{1}{c|}{$\neq$ chord$_2$}             \\
			\cline{1-3}
			\multicolumn{1}{r}{$=$ chord$_1$}    & \multicolumn{1}{|c|}{$E_{11} = \frac{R_1C_1}{N}$}   & \multicolumn{1}{c|}{$E_{12} = \frac{R_1C_2}{N}$}         \\
			\cline{1-3}
			\multicolumn{1}{r}{$\neq$ chord$_1$}   & \multicolumn{1}{|c|}{$E_{21} = \frac{R_2C_1}{N}$}  & \multicolumn{1}{c|}{$E_{22} = \frac{R_2C_2}{N}$}       \\
			\cline{1-3}
			&  &   \\
		\end{tabular}%
	\end{minipage}
	\caption{The general form of the $2\times2$ contingency table for bigrams, with observed frequencies and row and column marginals (left), and expected frequencies under the null hypothesis of independence (right).}
	\label{fig:contingency}%
\end{figure}%

The statistical analysis of contingency tables compares the observed frequencies in the left table against the expected frequencies in the right table under the null hypothesis that the rows and columns are statistically independent. The estimates in these tables can then be used to calculate a large number of AMs. Although some of these measures have become de-facto standards, such as log-likelihood in computational linguistics and mutual information in computational lexicography, there is no ideal AM. \citet{Pecina2005} identified 57 measures in current usage, and new measures and variants are constantly being invented \citep[32]{Evert2008}. To complicate matters, AMs are typically defined only for bigrams, though methods for extending AMs have recently been suggested \citep{McInnes2004, Petrovic2010}. Thus, we have adapted and extended four of the most well-known AMs for this study: pointwise mutual information, the Dice coefficient, and the chi-squared and log-likelihood statistics.  



\subsection{Pointwise Mutual Information}

According to \citet{Evert2008}, the most intuitive way to relate observed and estimated frequencies for a given $n$-gram type, $\mathcal{T}$, is to use the ratio $O_{11}/E_{11}$, with values larger than 1 indicating a positive statistical association. However, since the value of $O_{11}/E_{11}$ can become extremely high for large corpora, it is more convenient to measure association on a (base-2) logarithmic scale, yielding a statistic known as \textit{pointwise mutual information}, or \textit{pMI} \citep{Church1990}. \textit{pMI} can easily be extended to accommodate types of any cardinality $n$ using the following equation:

\begin{equation}\label{pmi}
pMI(\mathcal{T}) = \log_2\frac{P(\text{chord}_1 \ldots \text{chord}_n)}{\prod_{i=1}^nP(\text{chord}_i)}
\end{equation}
The numerator estimates $O_{11}$ for $\mathcal{T}$, and the denominator refers to the joint probability of the constituent chord members in $\mathcal{T}$, $E_{11}$.  In information theory, this value can be interpreted as the number of bits of shared information between $n$ events. 

Unfortunately, \textit{pMI} is known to favor rare types in the $n$-gram list. As a result, even a single co-occurrence of two or more chord members can result in a fairly high association score. In order to counterbalance this low-frequency bias, several heuristic modifications have been proposed. The first of these modifications, called $pMI_{\text{local}}$, weights the $pMI$ estimate by the pattern's observed probability \citep{Evert2008}.
 
\begin{equation}\label{pmi-local}
pMI_{\text{local}}(\mathcal{T}) = P(\text{chord}_1 \ldots \text{chord}_n) \times \log_2\frac{P(\text{chord}_1 \ldots \text{chord}_n)}{\prod_{i=1}^nP(\text{chord}_i)}
\end{equation}
The goal of $pMI_{\text{local}}$ is to scale $\mathcal{T}$ by its observed probability, thereby privileging types with larger observed probabilities in the final list. 

The second modification employed here adopts a similar approach to Equation \ref{pmi-local} by weighting $pMI$ by an observed probability, but in this case the weighting parameter is a coverage statistic that measures the proportion of compositions in the corpus that contain $\mathcal{T}$. The assumption here is that particularly characteristic voice-leading patterns like the MRDCC may only feature a few instances \textit{within} any given composition, but will nevertheless cover a large number of compositions across the corpus.
\begin{equation}\label{coverage}
coverage(\mathcal{T}) = \frac{|m:\mathcal{T} \in m|}{|C|}
\end{equation}
Here, $|m:\tau \in m|$ refers to the total number of compositions in the corpus $C$ that contain $\mathcal{T}$ at least once. This coverage statistic then serves as a scaling factor in place of the observed probability estimate from Equation \ref{pmi-local}.

\begin{equation}\label{pmi-coverage}
pMI_{\text{coverage}}(\mathcal{T}) =  coverage(\mathcal{T}) \times \log_2\frac{P(\text{chord}_1 \ldots \text{chord}_n)}{\prod_{i=1}^nP(\text{chord}_i)}
\end{equation}
Thus, $pMI_\text{coverage}$ privileges types that appear in a greater proportion of compositions in the corpus.

\subsection{Dice}

Another common measure for ranking collocations is the Dice coefficient \citep{Dice1945}. Rather than comparing an observed frequency to an expected frequency that assumes independence, as is the case with $pMI$, the Dice coefficient focuses on cases of very strong positive association. As a result, it tends to privilege relatively rigid expressions in natural language corpora like \textit{New York Stock Exchange}, and so has been adopted by several text analysis software tools for the discovery of fixed multi-word units \citep{Smadja:1993, Kilgarriff2004}. 

\begin{equation}\label{dice}
Dice(\mathcal{T}) = \frac{nf(\text{chord}_1 \ldots \text{chord}_n)}{\sum_{i=1}^nf(\text{chord}_i)}
\end{equation}
Here, $f()$ calculates the frequencies for $\mathcal{T}$ and its constituent chord members, and $n$ refers to the cardinality of $\mathcal{T}$. Types with strong positive associations will receive a value close to 1.

\subsection{Chi-squared}
$pMI$ and Dice are effect-size AMs because they attempt to quantify the statistical strength of the association between events. As a result, they generally fail to take sampling variation into account. Thus, statistical significance measures are often proposed as alternatives because they rank each $n$-gram type according to the amount of statistical evidence provided by a sample against the null hypothesis of independence. The most appropriate significance test is generally assumed to be Fisher's exact test because it does not rely on approximations that may be invalid for low-frequency data, but \citet{Evert2008} found that the chi-square ($\chi^2$) and log-likelihood ($G^2$) tests provide excellent approximations to Fisher's test and are much easier to implement, so we have selected those measures here.

To calculate the statistical association between events when $n>2$, $pMI$ and Dice rely on a relatively straightforward extension that \citet{Petrovic2010} refer to as $G_0$, the base-case extension. In short, $G_0$ is specific to each AM and treats all events in an $n$-gram equally. However, applying $G_0$ to the chi-squared and log-likelihood statistics is computationally expensive due to the \textit{n}-dimensionality of the contingency tables on which these statistics depend.\footnote{For a worked example of contingency tables in pattern discovery, see \citet{Sears2019}.} Thus, we have elected to apply an extension that \citet{daSilva1999} have called \textit{fair dispersion point normalization}, labeled as $G_5$ in the extension pattern list described by \citet{Petrovic2010}.   

\begin{equation}\label{G5}
G_5(g, \text{chord}_1 \ldots \text{chord}_n) = \frac{1}{n-1}\sum_{i=1}^{n-1}g(\text{chord}_1 \ldots \text{chord}_i, \text{chord}_{i+1} \ldots \text{chord}_n)
\end{equation}
$G_5$ divides the $n$-gram into all possible two-component sub-sequences that take every $n$-gram member into account. For example, in the $n$-gram \textit{New York Stock Exchange}, $G_5$ would compute the strength of \textit{New} and \textit{York Stock Exchange}, \textit{New York} and \textit{Stock Exchange}, and \textit{New York Stock} and \textit{Exchange}. It then averages these values to obtain the final association score.

\begin{equation}\label{chi-squared}
\chi^2(\mathcal{T}) = \sum_{ij}\frac{(O_{ij} - E_{ij})^2}{E_{ij}}
\end{equation}
The chi-squared statistic sums the squared error between $O_{ij}$ and $E_{ij}$ for all cells of the contingency table. It is often the preferred test for independence in contingency tables because it gives an excellent approximation to the limiting $\chi^2$ distribution for small samples \citep{Agresti2002}. However, it has also been shown to overestimate significance for co-occurrence data in natural language corpora \citep[33]{Evert2008}.

\subsection{Log-likelihood}

Unlike the $\chi^2$ statistic, the log-likelihood measure generally produces better approximations of Fisher's exact test \citep{Dunning1993}, and variants of $G^2$ have also been applied in pattern discovery algorithms for symbolic music corpora \citep{Collins2016}.  

\begin{equation}\label{G^2}
G^2(\mathcal{T}) = 2\sum_{ij}O_{ij}\log\frac{O_{ij}}{E_{ij}}
\end{equation}
Following \citet{Evert2008}, in cases where the logarithm is undefined due to empty cells in the contingency table, the term evaluates to zero and can be omitted from the summation.

\section{Methods}\label{sec:methods}

\begin{figure}[t!]
	\centering{}
	\fontsize{8pt}{10pt}\selectfont
	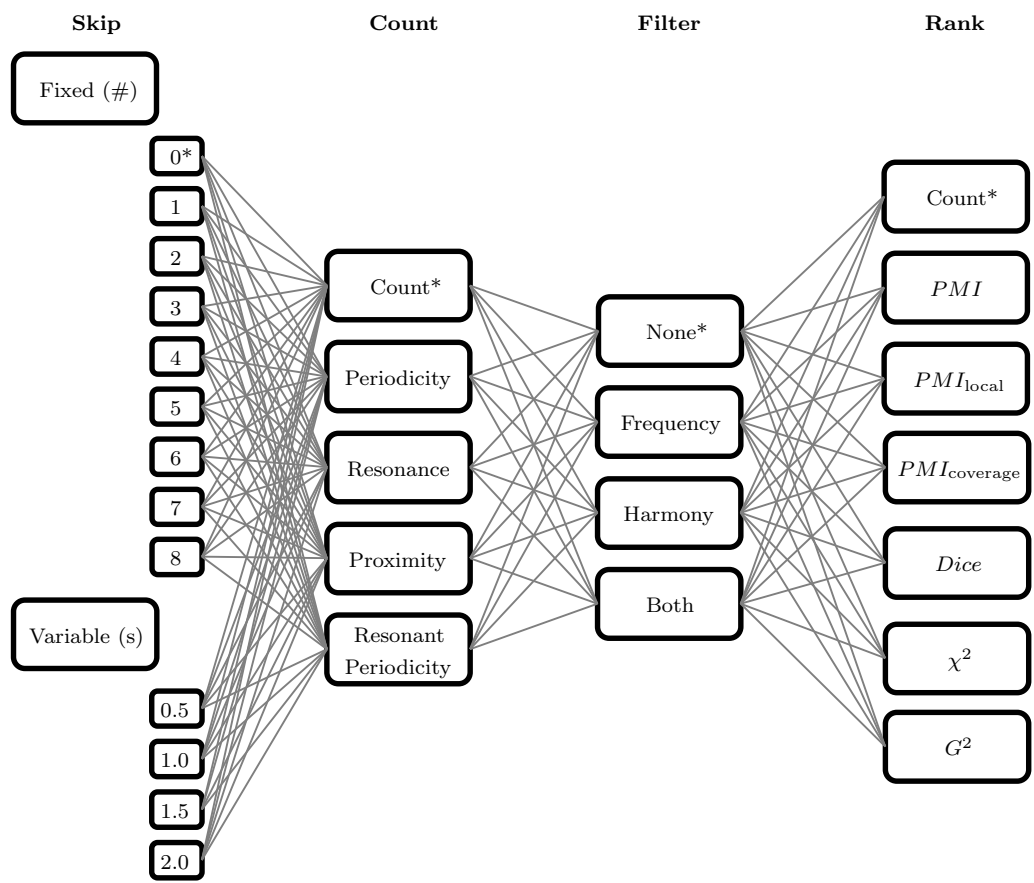
	\caption{The model configuration pipeline. Levels marked with an asterisk serve as the baseline comparison for that stage of the pipeline.}
	\label{fig:pipeline}
\end{figure}

Figure \ref{fig:pipeline} presents the model configuration pipeline. Altogether, the pipeline includes methods for defining (9 fixed-skip and 4 variable-skip levels), counting (5 levels), filtering (4 levels), and ranking (7 levels) $n$-gram types, yielding 1820 total model configurations.  

\subsection{Corpora}\label{sec:corpora}

Shown in Table \ref{tab:corpus}, this study includes 275 compositions from four data sets of Western classical music. Each data set features symbolic representations of both the notated score (e.g., metric position, rhythmic duration, pitch, etc.) and a recorded expressive performance (e.g., event onset time and duration in seconds, velocity, etc.), and includes both string quartet and piano textures. 

\begin{table}[t!]
	\centering
	\tbl{Data sets and descriptive statistics for the corpus.}
	{\begin{tabular}{l S[table-format=3.0] S[table-format=6.0] S[table-format=3.0]}
			\toprule
			Composer (Performer(s)) & \textit{N}$_{\text{compositions}}$ & {\textit{N}$_{\text{chords}}$} & {\textit{N}$_{\text{tokens}>3}$}\\
			\midrule
			Haydn (Kod\'{a}ly) & 50 & 73704 & 0\\
			Mozart (Batik) & 39 & 63418 & 969\\
			Beethoven (Zeilinger) & 30 & 42157 & 910\\
			Chopin (Magaloff) & 156 & 147871 & 3666\\
			\addlinespace[.2cm]
			\multicolumn{1}{r}{\textit{Total}} & 275 & 327150 & 5545\\
			\bottomrule
	\end{tabular}}%
	\tabnote{\textit{Note}. \textit{N}$_{\text{tokens}>3}$ denotes \textit{n}-gram tokens that initially consisted of more than three distinct interval classes.}
	\label{tab:corpus}%
\end{table}%

The Haydn/Kod\'{a}ly data set consists of 50 Haydn string quartet movements encoded in MIDI format \citep{Sears2016}. The data were manually aligned at the downbeat level to recorded performances by the Kod\'{a}ly Quartet, and then  the onset time for each chord event in the symbolic representation was estimated using linear interpolation. For the remaining data sets, performances were recorded on a B\"{o}sendorfer SE 290 computer-controlled piano, which is equipped with sensors on the keys and hammers to measure the timing and dynamics of each note \citep{Widmer:2003}. The Mozart/Batik data set consists of 13 complete Mozart piano sonatas (39 movements) encoded in MATCH format and performed by Roland Batik \citep{Widmer:2001}. The Beethoven/Zeilinger data set consists of 9 complete Beethoven piano sonatas (30 movements) encoded in MusicXML format and performed by Clemens Zeilinger \citep{Flossmann:2010}. Finally, the Chopin/Magaloff data set consists of 156 Chopin piano works that were performed by Nikita Magaloff \citep{Flossmann:2010,Flossmann:2010b}. 

To derive chord events from the corpus, we performed a full expansion of each piece, which produced 179,279 distinct onsets. However, some of the onsets included more than three vertical interval classes. Since the VLT scheme only permits up to three interval classes, $S$, above the bass, we replaced any chord containing more than three interval classes with the most common chord featuring the closest maximal subset of interval classes (e.g., $\langle4,7,10,11\rangle$ would likely reduce to $\langle4,7,10\rangle$). The most common chord was measured either from (1) the immediate surrounding context (i.e., $\pm$ 5 chords); (2) the entire piece; or finally (3) the entire corpus. This process replaced 5545 chord onsets, or less than 2\% of all chord events in the corpus.

\subsection{Evaluation}

One standard approach for evaluating query results is to manually annotate all candidates in an $n$-best list as either \textit{true positives} (i.e., expressions deemed to be important by the expert annotator) or \textit{false positives}. These annotations are then used to calculate the precision of the $n$-best list \citep[35]{Evert2008}, which in turn may be used to compare configurations representing each level of a given stage of the model pipeline (e.g., fixed: 0 skips vs. 4 skips). These methods are generally quite common in corpus linguistics, where the identification of meaningful collocations is assumed to be straightforward. In our case, however, we have elected to treat the MRDCC as an exemplar for the sorts of patterns that we \textit{assume} should be relevant for this corpus. Thus, we will privilege model configurations that produce the highest possible rank in the final (count- or AM-) sorted list for the MRDCC using an information retrieval metric called \textit{mean reciprocal rank} (MRR) \citep{Vorhees2000}. 

\begin{equation} \label{MRR}
MRR = \frac{1}{|Q|}\sum_{i=1}^{|Q|}\frac{1}{\text{rank}_i}
\end{equation}
Here, rank$_i$ refers to the rank of the MRDCC in the sorted list, and $Q$ refers to the total number of model configurations for a given level in the pipeline. Thus, the MRR is calculated from the MRDCC ranks corresponding to all lists that include that level in the pipeline (e.g., all configurations that include \textit{Skip=2}). Higher values of MRR indicate higher ranks in the lists featuring that level, with 1.0 indicating a perfect (i.e., top) rank in all lists. 

In the analyses that follow, we compare the MRR estimated for the best performing level from each stage of the pipeline to the MRR from the corresponding baseline level using an independent, two-samples $t$-test. The following levels serve as baseline levels: \textit{Skip} = 0; \textit{Count} = Count; \textit{Filter} = None; and \textit{Rank} = Count. To minimize the risk associated with calculating multiple comparisons, each test was corrected with Bonferroni adjustment, which divides the significance criterion by the number of planned comparisons. Finally, we also report Cohen's $d$ to estimate the size of the effect for each comparison.    

%
%

\section{Results}\label{sec:results}

Table \ref{tab:comparisons} presents the pairwise comparisons between the best-performing levels from each stage of the pipeline and their corresponding baseline levels. Given the potential differences in texture between these data sets, we considered each data set separately, as well as the union of all sets (denoted as \textit{All} in Table \ref{tab:comparisons}). Positive coefficients for a given comparison indicate that configurations from the best-performing level received higher MRR estimates than the baseline level, thereby ranking the MRDCC higher in their corresponding $n$-gram lists. 

\begin{table}[t!]
	\centering
	\tbl{Pairwise comparisons for the best-performing and baseline levels from each stage of the model pipeline for all data sets.}
	{\begin{tabular}{
				l S[group-separator=\text{.}, parse-numbers=false] S[group-separator=false,add-integer-zero=false, table-format=0.3] S[table-format=1.3,table-space-text-pre={***}, table-space-text-post={***}] S[table-format=4.0] S[add-integer-zero=false, table-format=0.3]}
			\toprule
			& \multicolumn{1}{c}{Comparison} & \multicolumn{1}{c}{$\Delta_{MRR}$}   & \multicolumn{1}{c}{$t$} & \multicolumn{1}{c}{$df$} & \multicolumn{1}{c}{$d$} \\
			\midrule
			\textit{Haydn (Kod\'{a}ly)} &       &       &       &       &  \\
			\quad Skip & \text{Variable vs}. \text{ Fixed} & -.019  & -4.499*** & 1818  & -.229 \\
			\qquad Fixed & \text{4 skips vs}. \text{ 0 skips} & .050 & 5.309*** & 278   & .635 \\
			\qquad Variable & \text{2s vs}. \text{ 0 skips} & .023 & 4.520*** & 278   & .540 \\
			\quad Count & \text{Periodicity vs}. \text{ Count} & -.005  & -0.724 & 726  & -.054 \\
			\quad Filter & \text{Harmony vs}. \text{ None} & .056 & 9.667*** & 908  & .641 \\
			\quad Rank & pMI\text{\num{_{coverage}} vs}. \text{ Counts} & .093 & 7.992*** & 518   & .701 \\
			\addlinespace[.2cm]
			\textit{Mozart (Batik)} &       &       &       &       &  \\
			\quad Skip & \text{Variable vs}. \text{ Fixed} & -7.080{\text{e}$^{-5}$} & -7.216*** & 1818  & -.367 \\
			\qquad Fixed & \text{6 skips vs}. \text{ 0 skips} & 1.564{\text{e}$^{-4}$} & 7.009*** & 278   & .838 \\
			\qquad Variable & \multicolumn{1}{c}{NA}    &     &     &   &  \\
			\quad Count & \text{Periodicity vs}. \text{ Count} & -4.470{\text{e}$^{-6}$} & -0.265 & 726  & -.020 \\
			\quad Filter & \text{Both vs}. \text{ None} & 8.891{\text{e}$^{-5}$} & 6.111*** & 908  & .405 \\
			\quad Rank & pMI\text{\num{_{coverage}} vs}. \text{ Counts} & 1.309{\text{e}$^{-4}$} & 4.368*** & 518   & .383 \\
			\addlinespace[.2cm]
			\textit{Beethoven (Zeilinger)} &       &       &       &       &  \\
			\quad Skip & \text{Variable vs}. \text{ Fixed} & -.005  & -3.914** & 1818  & -.199 \\
			\qquad Fixed & \text{2 skips vs}. \text{ 0 skips} & .005 & 1.344 & 278   & .161 \\
			\qquad Variable & \text{1s vs}. \text{ 0 skips} & -.002  & -.845 & 278   & -.101 \\
			\quad Count & \text{Resonance vs}. \text{ Count} & .002 & 1.390 & 726  & .103 \\
			\quad Filter & \text{Both vs}. \text{ None} & .012 & 6.692*** & 908  & .459 \\
			\quad Rank & pMI\text{\num{_{coverage}} vs}. \text{ Counts} & .026 & 7.778*** & 518   & .682 \\
			\addlinespace[.2cm]
			\textit{Chopin (Magaloff)} &       &       &       &       &  \\
			\quad Skip & \text{Variable vs}. \text{ Fixed} & -.0003  & -1.455 & 1818  & -.074 \\
			\qquad Fixed & \text{8 skips vs}. \text{ 0 skips} & .002 & 3.385* & 278   & .405 \\
			\qquad Variable & \text{2s vs}. \text{ 0 skips} & .002  & 3.321* & 278   & .397 \\
			\quad Count & \text{Resonant Periodicity vs}. \text{ Count} & .0002 & .555 & 726  & .041 \\
			\quad Filter & \text{Harmony vs}. \text{ None} & .001 & 4.763*** & 908  & .316 \\
			\quad Rank & pMI\text{\num{_{coverage}} vs}. \text{ Counts} & .004 & 6.965*** & 518   & .611 \\
			\addlinespace[.2cm]
			\textit{All} &       &       &       &       &  \\
			\quad Skip & \text{Variable vs}. \text{ Fixed} & -.010  & -3.289* & 1818  & -.167 \\
			\qquad Fixed & \text{5 skips vs}. \text{ 0 skips} & .033 & 4.517*** & 278   & .540 \\
			\qquad Variable & \text{2s vs}. \text{ 0 skips} & .021 & 3.795** & 278   & .454 \\
			\quad Count & \text{Periodicity vs}. \text{ Count} & -.002  & -.553  & 726  & -.041 \\
			\quad Filter & \text{Harmony vs}. \text{ None} & .037 & 9.569*** & 908  & .634 \\
			\quad Rank & pMI\text{\num{_{coverage}} vs}. \text{ Counts} & .072 & 9.083*** & 518   & .797 \\
			\bottomrule
	\end{tabular}}%
	\tabnote{\textit{Note.} $\Delta_{MRR}$ refers to the average difference in reciprocal rank, $t$ is an independent two-sample $t$-test, $df$ denotes the degrees-of-freedom, and $d$ refers to Cohen's $d$. All $p$-values are corrected with Bonferroni adjustment. *$p < .05$; **$p < .01$; ***$p < .001$.}
	\label{tab:comparisons}%
\end{table}%

\begin{figure}[p!]
	\centering{}
	\fontsize{8pt}{10pt}\selectfont
	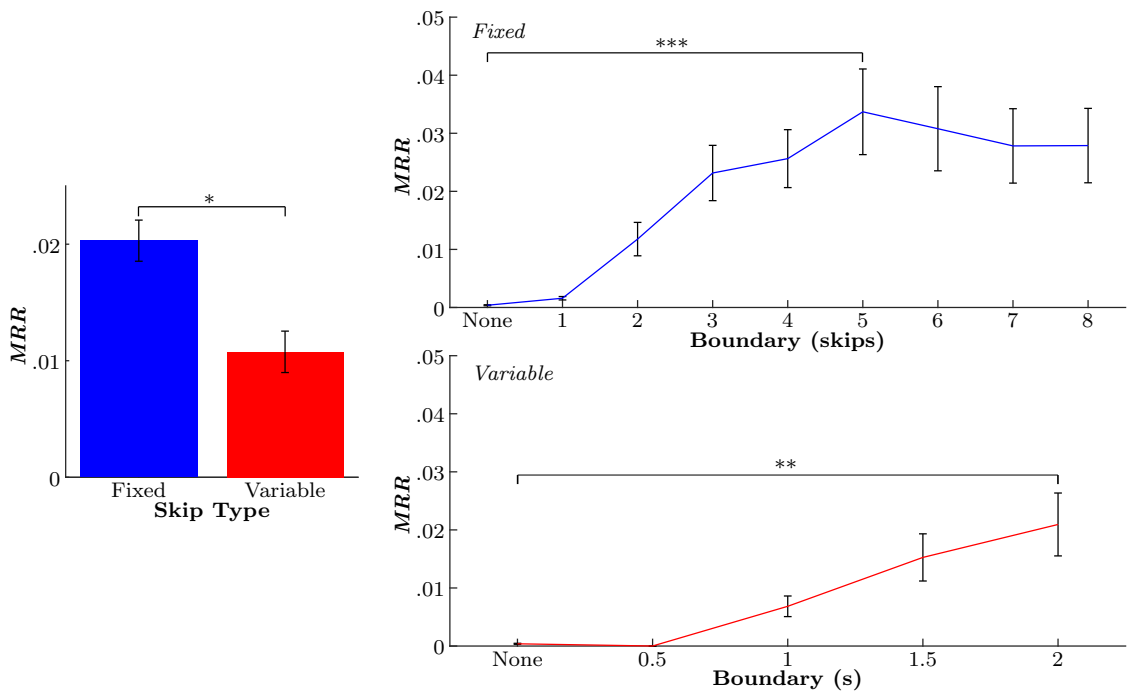
	\caption{Left: Bar plot of the MRR estimates across all fixed (blue) and variable (red) model configurations. Right: Line plots of the mean RR estimates for all fixed and variable skips. Error bars represent $\pm 2$ standard errors. *$p < .05$; **$p < .01$; ***$p < .001$.}
	\label{fig:RR_fig1}
\end{figure}

Figure \ref{fig:RR_fig1} presents bar and line plots of the MRR estimates across all fixed and variable model configurations for the entire corpus (i.e., \textit{All}).  Across all data sets, the MRR increased incrementally from no fixed skips (i.e., contiguous $n$-grams) to 5 fixed skips. A similar increase occurred for the variable-skip method, with the MRR estimate continuing to increase even for a variable-skip interval of 2s. The fixed skip method outperformed the variable-skip method overall, however, suggesting that the time-course of the MRDCC is itself quite variable, and thus may be more difficult to identify using a specified temporal interval. The variable-skip method also dramatically increases the number of tokens in the final count distribution, particularly for fast-tempo compositions, which may be an additional contributing factor to the cadence's reduced final rank. However, it is worth noting that effect sizes for the fixed vs. variable comparison decreased significantly for the Beethoven/Zeilinger and Chopin/Magaloff data sets, suggesting that the increased number of tokens produced by the variable-skip method may be useful for more dense, complex textures.  

The number of fixed skips in the best-performing level varied across all data sets (Haydn/Kod\'{a}ly: 4 skips; Mozart/Batik: 6 skips; Beethoven/Zeilinger: 2 skips; Chopin/Magaloff: 8 skips). This finding likely reflects differences in the number of MRDCC tokens that were identified using the skip-gram approach. In the Mozart/Batik data set, for example, the variable-skip method failed to identify even a single instance of the compound cadence (see `NA' in Table \ref{tab:comparisons}). As a result, changes in MRR across all model configurations were extraordinarily small in the Mozart/Batik data set. This result likely reflects the presence of accompanimental textures in Mozart's keyboard style that prevented the skip-gram method from identifying each harmony of the MRDCC in the expanded encoding.

\begin{figure}[p!]
	\centering{}
	\fontsize{8pt}{10pt}\selectfont
	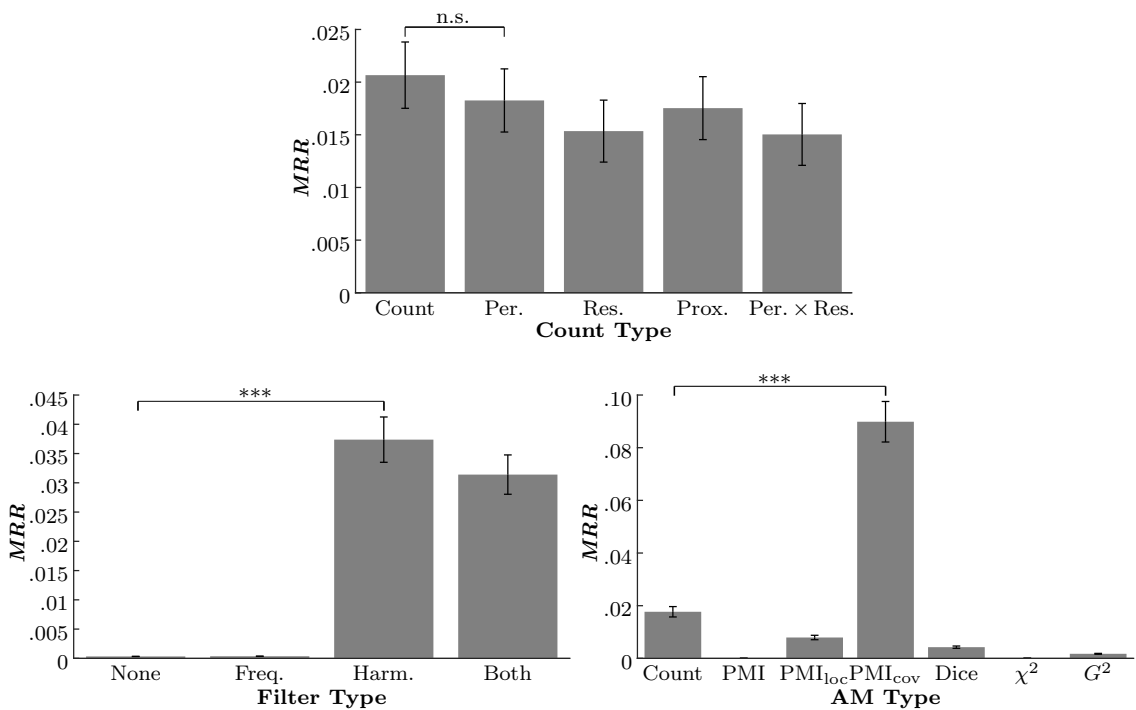
	\caption{Bar plots of the MRR estimates across all model configurations for count type, filter type, and AM type. pMI$_{\text{loc}}$ and pMI$_{\text{cov}}$ refer to the local- and coverage-weighted pMI measures, respectively. Error bars represent $\pm 2$ standard errors. *$p < .05$; **$p < .01$; ***$p < .001$.}
	\label{fig:RR_fig2}
\end{figure}

Figure \ref{fig:RR_fig2} presents bar plots of the MRR estimates across all data sets for count type, filter type, and AM-rank type. Overall, the periodicity function received the highest MRR estimates of the weighted count measures for three of the five data sets, but none of the weighted measures significantly increased the rank of the cadence relative to an unweighted count. For the Beethoven/Zeilinger and Chopin/Magaloff data sets, the resonance and resonant periodicity models outperformed the baseline level, but these differences were not significant. Thus, weighting measures based on recorded performances of the score may improve model performance for dense textures, but not significantly so. 

Of the filter types, the harmony filter yielded the greatest improvement in model performance in the Haydn/Kod\'{a}ly data set, the Chopin/Magaloff data set, and across all data sets. Filtering based on frequency and harmony (Both) produced the best results for the Mozart/Batik and Beethoven/Zeilinger data sets, which should not be surprising given the greater difficulty associated with finding the MRDCC in these data sets. Thus, filtering plays an important role in the model pipeline, with the harmony filter clearly producing the greatest increase in MRR. 

Finally, several AMs ranked the MRDCC in the top 100 of the final $n$-gram list. The $pMI$ statistic generally performed worst of the AMs included here. In this case, the equation's well-known frequency bias resulted in an $n$-best list comprised entirely of patterns with extraordinarily low counts, which is one reason corpus linguists apply a frequency filter before ranking-sorting the list. Including scaling heuristics to offset this bias increased the MRR significantly for $pMI_{\text{local}}$, which weights the $pMI$ estimate by its observed probability, and for $pMI_{\text{coverage}}$, which weights $pMI$ by a coverage statistic representing the proportion of compositions featuring that $n$-gram type. Of the two AMs, $pMI_{\text{coverage}}$ also significantly outperformed the baseline level, which ranked each $n$-gram type by its count. It is also noteworthy that the Dice and $G^2$ statistics ranked the MRDCC in the top 100, which indicates that (1) the constituent members of the MRDCC feature particularly strong positive statistical associations, and (2) there is sufficient statistical evidence to reject the null hypothesis of independence.     

\begin{figure}[t!]
	\centering{}
	\fontsize{8pt}{10pt}\selectfont
	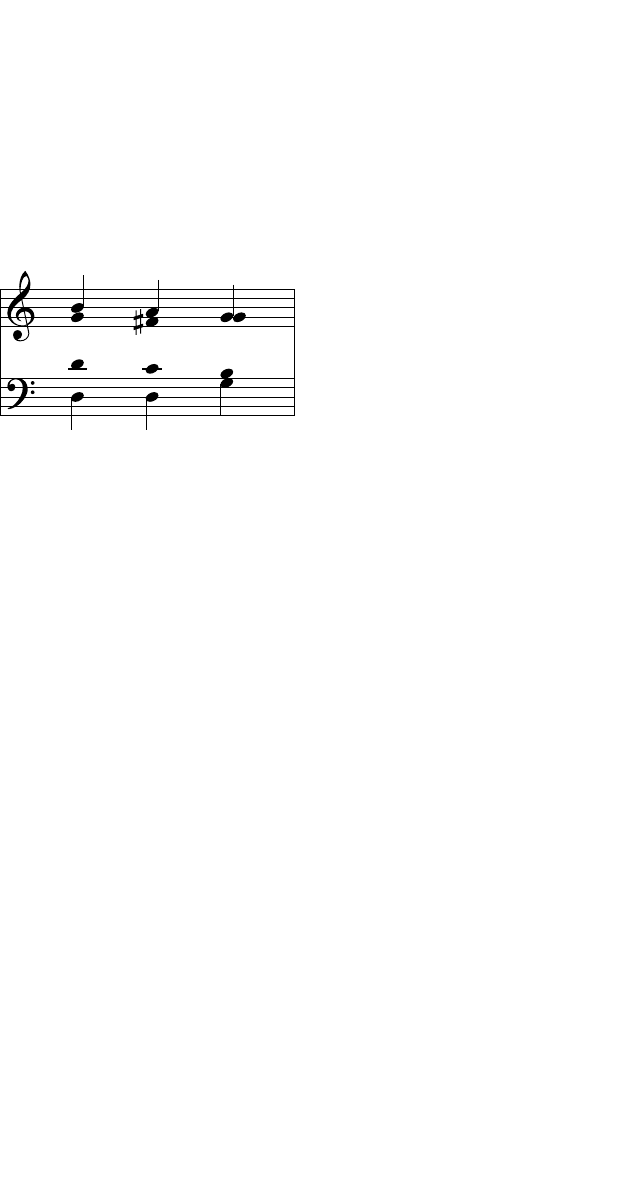
	\caption{Top ten trigram types notated in the key of G-major. Outer voices are notated with stems because the VLT scheme specifies the contrapuntal organization of these voices. Melodic scale-degrees and Roman numeral interpretations are provided above and below.}
	\label{fig:top10}
\end{figure}

On the basis of these findings, the optimal model configuration consists of the following parameters: \textit{Type} = Fixed; \textit{Skip} = Five; \textit{Weight} = Count; \textit{Filter} = Harmony; \textit{Rank} = $pMI_{\text{coverage}}$. So what sorts of patterns will emerge at the top of the $n$-gram list? Table \ref{tab:top10trigrams} provides the top ten trigram types identified by this model configuration. To facilitate interpretation, we have included the VLT encoding for each type, along with Roman numeral and melodic scale-degree notation for the most likely tonal harmonic progression it represents. Each progression is further categorized using the three fundamental progressions of harmony described by \citet{Caplin1998}: \textit{prolongational} progressions, which sustain an individual harmony through other (subordinate) harmonies (e.g., I--V$^4_3$--I$^6$); \textit{sequential} progressions, which project a contrapuntal pattern independent of harmonic functionality (e.g., IV$^6$--iii$^6$--ii$^6$); and \textit{cadential} progressions, which confirm a tonal center (e.g., V$^6_4$--V$^7$--I).\footnote{In Caplin's form-functional approach to tonal harmony, prolongational progressions tend to initiate a larger phrase-structural process, such as a phrase or theme (\textit{beginning}), sequential progressions continue that process (\textit{middle}), and cadential progressions (attempt to) conclude it (\textit{end}).} Finally, Figure \ref{fig:top10} realizes the top 10 trigram types using Western notation. Outer voices are notated with stems to remind readers that the VLT scheme specifies the contrapuntal organization of these voices.\footnote{The contrapuntal organization of the inner voices in Figure \ref{fig:top10} is thus an interpretation on our part.} 

\begin{table}[t!]
	\centering
	\tbl{Top ten trigram types identified by the optimal model configuration.}
	{\begin{tabular} 
			{c c l c l c l r S[output-decimal-marker={},parse-numbers=false] S[output-decimal-marker={},parse-numbers=false] S[output-decimal-marker={},parse-numbers=false] l l}
			\toprule
			$pMI_{\text{cov.}}$ & & \multicolumn{5}{c}{\textit{VLT} ($S_1,I_1,S_2,I_2,S_3$)}      &       & \multicolumn{3}{c}{\textit{RNA}} & & \multicolumn{1}{c}{\textit{Type}} \\
			\cmidrule{1-1} \cmidrule{3-7} \cmidrule{9-11} \cmidrule{13-13}
			\multirow{2}{*}{1.796}   &  & \multirow{2}{*}{$<$4,9*,10$>$} & \multirow{2}{*}{0}     & \multirow{2}{*}{$<$4,7*,10$>$} & \multirow{2}{*}{5}     & \multirow{2}{*}{$<$4,$\perp$,$\perp$$>$} &       & \hat{3} & \hat{2} & \hat{1} & & \multirow{2}{*}{cadential} \\
			& & & & & & & & \text{V}.^7 & \text{V}.^7    & \text{I}. & & \\ \addlinespace[1pt] \hdashline \addlinespace[3pt]
			\multirow{2}{*}{1.483}   &  & \multirow{2}{*}{$<$5*,9,$\perp$$>$} & \multirow{2}{*}{0}     & \multirow{2}{*}{$<$4,7*,10$>$} & \multirow{2}{*}{5}     & \multirow{2}{*}{$<$4,$\perp$,$\perp$$>$} &       & \hat{1} & \hat{2} & \hat{1} & & \multirow{2}{*}{cadential}\\
			& & & & & & & & \text{V}.(^6_4) & \text{V}.^7    & \text{I}. & & \\  \addlinespace[1pt] \hdashline \addlinespace[3pt]
			\multirow{2}{*}{1.069}   &  & \multirow{2}{*}{$<$5,9*,$\perp$$>$} & \multirow{2}{*}{0}     & \multirow{2}{*}{$<$4,7*,10$>$} & \multirow{2}{*}{5}     & \multirow{2}{*}{$<$4,$\perp$,$\perp$$>$} &       & \hat{3} & \hat{2} & \hat{1} & & \multirow{2}{*}{cadential}\\
			& & & & & & & & \text{V}.(^6_4) & \text{V}.^7    & \text{I}. & & \\ \addlinespace[1pt] \hdashline \addlinespace[3pt]
			\multirow{2}{*}{0.931}   &  & \multirow{2}{*}{$<$3,8*,$\perp$$>$} & \multirow{2}{*}{0}     & \multirow{2}{*}{$<$3,6,$\perp$$>$} & \multirow{2}{*}{1}     & \multirow{2}{*}{$<$4*,$\perp$,$\perp$$>$} &       & \hat{5} & \hat{4} & \hat{3} & & \multirow{2}{*}{prolongational}\\
			& & & & & & & & \text{V}.^6 & \text{V}.^6_5    & \text{I}. & & \\ \addlinespace[1pt] \hdashline \addlinespace[3pt]
			\multirow{2}{*}{0.834}   &  & \multirow{2}{*}{$<$4,7*,$\perp$$>$} & \multirow{2}{*}{5}     & \multirow{2}{*}{$<$4*,$\perp$,$\perp$$>$} & \multirow{2}{*}{7}     & \multirow{2}{*}{$<$4,7*,$\perp$$>$} &       & \hat{2} & \hat{3} & \hat{2} & & \multirow{2}{*}{prolongational}\\
			& & & & & & & & \text{V}. & \text{I}.    & \text{V}. & & \\ \addlinespace[1pt] \hdashline \addlinespace[3pt]
			\multirow{2}{*}{0.796}   &  & \multirow{2}{*}{$<$4,9*,$\perp$$>$} & \multirow{2}{*}{11}     & \multirow{2}{*}{$<$3,8*,$\perp$$>$} & \multirow{2}{*}{10}     & \multirow{2}{*}{$<$3,9*,$\perp$$>$} &       & \hat{2} & \hat{1} & \hat{7} & & \multirow{2}{*}{sequential}\\
			& & & & & & & & \text{ii}.^6 & \text{I}.^6    & \text{vii}.^6 & & \\ \addlinespace[1pt] \hdashline \addlinespace[3pt]
			\multirow{2}{*}{0.795}   &  & \multirow{2}{*}{$<$4,9*,$\perp$$>$} & \multirow{2}{*}{10}     & \multirow{2}{*}{$<$4,9*,$\perp$$>$} & \multirow{2}{*}{11}     & \multirow{2}{*}{$<$3,8*,$\perp$$>$} &       & \hat{3} & \hat{2} & \hat{1} & & \multirow{2}{*}{sequential}\\
			& & & & & & & & \text{iii}.^6 & \text{ii}.^6    & \text{I}.^6 & &\\ \addlinespace[1pt] \hdashline \addlinespace[3pt]
			\multirow{2}{*}{0.784}   &  & \multirow{2}{*}{$<$4,9*,$\perp$$>$} & \multirow{2}{*}{2}     & \multirow{2}{*}{$<$5*,9,$\perp$$>$} & \multirow{2}{*}{0}     & \multirow{2}{*}{$<$4,7*,10$>$} &       & \hat{2} & \hat{1} & \hat{7} & & \multirow{2}{*}{cadential}\\
			& & & & & & & & \text{ii}.^6 & \text{V}.(^6_4)    & \text{V}.^7 & & \\ \addlinespace[1pt] \hdashline \addlinespace[3pt]
			\multirow{2}{*}{0.782}   &  & \multirow{2}{*}{$<$3,8*,$\perp$$>$} & \multirow{2}{*}{8}     & \multirow{2}{*}{$<$4,9*,$\perp$$>$} & \multirow{2}{*}{11}     & \multirow{2}{*}{$<$3,8*,$\perp$$>$} &       & \hat{1} & \hat{6} & \hat{5} & & \multirow{2}{*}{sequential ?}\\
			& & & & & & & & \text{I}.^6 & \text{vi}.^6    & \text{V}.^6 & &\\ \addlinespace[1pt] \hdashline \addlinespace[3pt]
			\multirow{2}{*}{0.781}   &  & \multirow{2}{*}{$<$5,9*,$\perp$$>$} & \multirow{2}{*}{0}     & \multirow{2}{*}{$<$4,7*,$\perp$$>$} & \multirow{2}{*}{5}     & \multirow{2}{*}{$<$0*,$\perp$,$\perp$$>$} &       & \hat{3} & \hat{2} & \hat{1} & & \multirow{2}{*}{cadential}\\
			& & & & & & & & \text{V}.(^6_4) & \text{V}.    & \text{I}. & &\\ 
			\bottomrule
	\end{tabular}}%
	\tabnote{\textit{Note.} Model parameters: \textit{Type} = Fixed; \textit{Skip} = Five; \textit{Weight} = Count; \textit{Filter} = Harmony; \textit{Rank} = $pMI_{\text{coverage}}$. \textit{VLT}: Numbers marked with an asterisk denote the interval class of the highest voice, $T$, above the bass. VLT members without an asterisk indicate that the highest voice doubles the bass at the unison or octave. \textit{RNA}: Numbers inside parentheses are figured bass symbols with the root in the bass.}
	\label{tab:top10trigrams}%
\end{table}%

Five of the top ten types represent cadential progressions. The highest ranked type in the $n$-gram list is a simple authentic cadence (V$^7$--I) supporting a Mi-Re-Do descent in the melody, with Mi serving as a non-chord tone. The second and third highest-ranked types represent two melodic variants of the compound cadence, with the initial cadential six-four supporting either Do or Mi. Similarly, the tenth-ranked VLT represents a harmonic variant of the MRDCC that omits the seventh of the penultimate dominant and the third of the final tonic. Finally, the eighth-ranked VLT is an antecedent progression from the compound cadence that includes the pre-dominant stage and supports a stepwise descent from Re to Ti.  

In addition to cadential progressions, Table \ref{tab:top10trigrams} includes prolongational and sequential progressions. The fourth-ranked type supports a stepwise descent from Sol to Mi through a first-inversion dominant-seventh chord. Note here that the harmony filter was intended to privilege types that feature harmonic change between adjacent members, but since patterns like the compound cadence prolong the dominant through multiple stages of the VLT, prolongational patterns like this one can also emerge. The fifth-ranked type similarly reflects the primacy of the tonic and dominant in the tonal system, in this case oscillating between V and I in root-position and Re and Mi in the melody. Finally, the types representing sequential progressions all feature six-three chords moving in parallel motion (called \textit{Fauxbourdon} technique). Since sequential progressions are less likely to be tied to an underlying tonal center, and the VLT scheme is invariant with respect to key, the Roman numeral annotations for these VLTs should also be interpreted more loosely. The presence of these types in the top 10 also results almost entirely from the inclusion of the Chopin/Magaloff data set, which tends to privilege parsimonious voice leading over functional harmonic progression. 

%




\section{Conclusion}\label{sec:conclusion}

To discover recurrent voice-leading patterns like the MRDCC, we extended the canonical $n$-gram approach---which divides a corpus into contiguous sequences of $n$ events---by using skip-grams, an alternative string-based method that includes sub-sequences in an $n$-gram list if their constituent members occur within a certain number of skips (\textit{fixed}), or a specified temporal interval (\textit{variable}). To that end, we compiled four data sets of Western tonal music consisting of symbolic encodings of both the notated score and a recorded performance, created a model pipeline for defining, counting, filtering, and ranking skip-grams, and ranked the position of the MRDCC in every possible model configuration. We found that the MRDCC receives a higher rank in the list when the pipeline employs fixed skips, filters the list by excluding $n$-gram types that do not reflect a genuine harmonic change between adjacent members, and ranks the remaining types using an extended statistical association measure like $pMI_{\text{coverage}}$. 

Despite the stylistic heterogeneity of the data sets included here, the MRDCC and its variants emerged at the top of an $n$-gram list that consisted of over 11 million tokens before filtering, and over 2 million tokens after filtering. Nevertheless, this discovery pipeline suffers from several limitations that should be addressed in future studies. First, we restricted the purview of vertical relations to temporally coincident (i.e., simultaneous) note events in the expanded encoding. This restriction seems reasonable for homorhythmic, chorale-like textures, but much less so for string quartets, piano sonatas, and the like, which often feature accompanimental textures that prolong harmonies over time (e.g., an Alberti bass pattern). \citet{Finkensiep2018} recently extended the skip-gram approach in a two-stage algorithm that identifies chords consisting of potentially non-coincident events within each notated measure before identifying progressions of those chords over time. However, the model does not store every possible voice-leading pattern due to the combinatoric complexity of the task, so future studies adopting a standard retrieval task will need to implement more efficient methods for search and storage.  

Second, none of the methods for weighting the counts using features of the recorded performance --- periodicity, resonance, proximity, and resonant periodicity --- significantly increased the rank of the MRDCC relative to an unweighted count. This finding was somewhat surprising in light of the claim that sampling events at regular temporal intervals improves pattern discovery \citep{Symons2012}. Our findings suggest that performance annotations could be irrelevant to pattern discovery tasks, which should benefit the community given the paucity of available performance data in musical corpus research. Nevertheless, future studies could examine whether symbolic, score-based features can improve model performance by privileging patterns whose constituent members appear in strong metric positions, feature long rhythmic durations, or include genuine changes of harmony. 

Third, by selecting the model configuration that optimized the rank of the MRDCC, the present approach biased the analysis pipeline towards a single pattern variant that appears across a large number of compositions. To discover other relevant intra- and inter-opus patterns from various genres and style periods (the Landini cadence of the Italian Trecento tradition, the double-plagal progression in 1970s rock music, etc.), future studies should optimize the ranks for a large number of patterns using an evaluation measure like average precision \citep{Petrovic2010}. To be sure, the performance of a given pattern discovery method will depend on the statistical properties of the pattern(s) the analyst hopes to study (e.g., whether its constituent members co-occur, whether it appears frequently within a given composition (intra-opus) or across several compositions (inter-opus), etc.). For example, by noting the probabilistic asymmetries between chords in a corpus of Haydn string quartets, \citet{Sears2019} was able to leverage asymmetric AMs to distinguish chord tones from non-chord tones in a harmonic reduction task. To be sure, future studies should attempt to combine an analysis pipeline like the one presented here with a method for assimilating pattern variants into their more general categories using statistical methods associated with similarity estimation and clustering, as was explored in \citet{Sears2016}.

Finally, corpus linguists have developed sophisticated taxonomies for collocations of various types in order to identify the appropriate AM for the expression at hand, but similar taxonomies have yet to be developed in the context of computational music analysis. Although the general definition of collocations as ``characteristic and frequently recurrent word combinations'' applies equally well to musical patterns like the MRDCC \citep[2]{Evert2008}, more rigid definitions of in computational linguistics generally do not apply. According to \citet{Manning1999}, collocations are typically defined according to three criteria: non-compositionality, non-substitutability, and non-modifiability. A collocation is non-compositional because its intended meaning is not a straightforward composition of its parts. An expression like \textit{Don't quit your day job}, for example,  is a wisecrack that has nothing to do with leaving one's profession. A collocation is also non-substitutable in that other words cannot be substituted for members of the collocation. Thus, \textit{Don't quit your day job} would lose its intended meaning --- or become unidiomatic to a native speaker --- if \textit{profession} replaced \textit{day job}. Finally, a collocation is also non-modifiable in that it cannot be freely modified with additional lexical material. \textit{Don't quit your amazing day job} would therefore surprise native speakers due to the rigid organization of the expression.   

Simply put, the MRDCC violates all of these criteria. It is compositional in that it characterizes the tonal system more broadly \citep{Casella1924}, substitutable in that the initial events of the pattern are often quite flexible, and modifiable in that the voice-leading scaffold rarely appears without extensive diminutions, as was seen in Figure \ref{fig:beethoven_complex}. Thus, by expanding the purview of possible voice-leading patterns to encompass sequences, cadences, and schemata of various sorts, the computational music analysis community could not only improve upon the current pattern discovery pipeline, but perhaps more importantly, develop a more sophisticated theory about the organizational principles that characterize recurrent patterns in music, cadential or otherwise.

\section*{Funding}
\addcontentsline{toc}{section}{Funding}

This research is supported by the European Research Council (ERC) under the EUs Horizon 2020 Framework Programme (ERC Grant Agreement number 670035, project ``Con Espressione'').
%
%
%
%

\bibliographystyle{tMAM}
\bibliography{main_bib}
\addcontentsline{toc}{section}{References}

\end{document}

%% file: beethoven_cad64_simple.pdf_tex
\begingroup%
  \makeatletter%
  \providecommand\color[2][]{%
    \errmessage{(Inkscape) Color is used for the text in Inkscape, but the package 'color.sty' is not loaded}%
    \renewcommand\color[2][]{}%
  }%
  \providecommand\transparent[1]{%
    \errmessage{(Inkscape) Transparency is used (non-zero) for the text in Inkscape, but the package 'transparent.sty' is not loaded}%
    \renewcommand\transparent[1]{}%
  }%
  \providecommand\rotatebox[2]{#2}%
  \ifx\svgwidth\undefined%
    \setlength{\unitlength}{173.71022049bp}%
    \ifx\svgscale\undefined%
      \relax%
    \else%
      \setlength{\unitlength}{\unitlength * \real{\svgscale}}%
    \fi%
  \else%
    \setlength{\unitlength}{\svgwidth}%
  \fi%
  \global\let\svgwidth\undefined%
  \global\let\svgscale\undefined%
  \makeatother%
  \begin{picture}(1,0.50788884)%
    \put(0.49042998,0.01242416){\color[rgb]{0,0,0}\makebox(0,0)[lb]{\smash{V($^6_4$)}}}%
    \put(0.72743373,0.01155391){\color[rgb]{0,0,0}\makebox(0,0)[lb]{\smash{I}}}%
    \put(0.59901387,0.01165351){\color[rgb]{0,0,0}\makebox(0,0)[lb]{\smash{V$^7$}}}%
    \put(0,0){\includegraphics[width=\unitlength,page=1]{beethoven_cad64_simple.pdf}}%
    \put(0.50907797,0.46327608){\color[rgb]{0,0,0}\makebox(0,0)[lb]{\smash{\tiny $\hat{3}$}}}%
    \put(0,0){\includegraphics[width=\unitlength,page=2]{beethoven_cad64_simple.pdf}}%
    \put(0.60163017,0.45257881){\color[rgb]{0,0,0}\makebox(0,0)[lb]{\smash{\tiny $\hat{2}$}}}%
    \put(0,0){\includegraphics[width=\unitlength,page=3]{beethoven_cad64_simple.pdf}}%
    \put(0.72412972,0.44164703){\color[rgb]{0,0,0}\makebox(0,0)[lb]{\smash{\tiny $\hat{1}$}}}%
    \put(0,0){\includegraphics[width=\unitlength,page=4]{beethoven_cad64_simple.pdf}}%
  \end{picture}%
\endgroup%

%% file: beethoven_cad64_complex.pdf_tex
\begingroup%
  \makeatletter%
  \providecommand\color[2][]{%
    \errmessage{(Inkscape) Color is used for the text in Inkscape, but the package 'color.sty' is not loaded}%
    \renewcommand\color[2][]{}%
  }%
  \providecommand\transparent[1]{%
    \errmessage{(Inkscape) Transparency is used (non-zero) for the text in Inkscape, but the package 'transparent.sty' is not loaded}%
    \renewcommand\transparent[1]{}%
  }%
  \providecommand\rotatebox[2]{#2}%
  \newcommand*\fsize{\dimexpr\f@size pt\relax}%
  \newcommand*\lineheight[1]{\fontsize{\fsize}{#1\fsize}\selectfont}%
  \ifx\svgwidth\undefined%
    \setlength{\unitlength}{215.5149139bp}%
    \ifx\svgscale\undefined%
      \relax%
    \else%
      \setlength{\unitlength}{\unitlength * \real{\svgscale}}%
    \fi%
  \else%
    \setlength{\unitlength}{\svgwidth}%
  \fi%
  \global\let\svgwidth\undefined%
  \global\let\svgscale\undefined%
  \makeatother%
  \begin{picture}(1,1.03297128)%
    \lineheight{1}%
    \setlength\tabcolsep{0pt}%
    \put(0,0){\includegraphics[width=\unitlength,page=1]{beethoven_cad64_complex.pdf}}%
    \put(0.1920846,0.01001417){\color[rgb]{0,0,0}\makebox(0,0)[lt]{\lineheight{1.25}\smash{\begin{tabular}[t]{l}V($^6_4$)\end{tabular}}}}%
    \put(0.63462444,0.00931267){\color[rgb]{0,0,0}\makebox(0,0)[lt]{\lineheight{1.25}\smash{\begin{tabular}[t]{l}I\end{tabular}}}}%
    \put(0.39746734,0.00939303){\color[rgb]{0,0,0}\makebox(0,0)[lt]{\lineheight{1.25}\smash{\begin{tabular}[t]{l}V$^7$\end{tabular}}}}%
    \put(0,0){\includegraphics[width=\unitlength,page=2]{beethoven_cad64_complex.pdf}}%
    \put(0.19954924,0.40746912){\color[rgb]{0,0,0}\makebox(0,0)[lt]{\lineheight{1.25}\smash{\begin{tabular}[t]{l}$\hat{3}$\end{tabular}}}}%
    \put(0,0){\includegraphics[width=\unitlength,page=3]{beethoven_cad64_complex.pdf}}%
    \put(0.50794444,0.39966704){\color[rgb]{0,0,0}\makebox(0,0)[lt]{\lineheight{1.25}\smash{\begin{tabular}[t]{l}$\hat{2}$\end{tabular}}}}%
    \put(0,0){\includegraphics[width=\unitlength,page=4]{beethoven_cad64_complex.pdf}}%
    \put(0.87007908,0.39209332){\color[rgb]{0,0,0}\makebox(0,0)[lt]{\lineheight{1.25}\smash{\begin{tabular}[t]{l}$\hat{1}$\end{tabular}}}}%
    \put(0,0){\includegraphics[width=\unitlength,page=5]{beethoven_cad64_complex.pdf}}%
    \put(0.06466488,0.41586567){\color[rgb]{0,0,0}\makebox(0,0)[lt]{\lineheight{1.25}\smash{\begin{tabular}[t]{l}(b)\end{tabular}}}}%
    \put(0.06465052,0.99710031){\color[rgb]{0,0,0}\makebox(0,0)[lt]{\lineheight{1.25}\smash{\begin{tabular}[t]{l}(a)\end{tabular}}}}%
    \put(0,0){\includegraphics[width=\unitlength,page=6]{beethoven_cad64_complex.pdf}}%
    \put(0.35975887,0.54223365){\color[rgb]{0,0,0}\makebox(0,0)[lt]{\lineheight{1.25}\smash{\begin{tabular}[t]{l}melodic events\end{tabular}}}}%
    \put(0,0){\includegraphics[width=\unitlength,page=7]{beethoven_cad64_complex.pdf}}%
    \put(0.35975887,0.49321342){\color[rgb]{0,0,0}\makebox(0,0)[lt]{\lineheight{1.25}\smash{\begin{tabular}[t]{l}harmonic events\end{tabular}}}}%
    \put(-0.3351236,0.49212538){\color[rgb]{0,0,0}\makebox(0,0)[lt]{\begin{minipage}{2.00967779\unitlength}\raggedright \end{minipage}}}%
  \end{picture}%
\endgroup%

%% file: skip_example_v2.pdf_tex
\begingroup%
  \makeatletter%
  \providecommand\color[2][]{%
    \errmessage{(Inkscape) Color is used for the text in Inkscape, but the package 'color.sty' is not loaded}%
    \renewcommand\color[2][]{}%
  }%
  \providecommand\transparent[1]{%
    \errmessage{(Inkscape) Transparency is used (non-zero) for the text in Inkscape, but the package 'transparent.sty' is not loaded}%
    \renewcommand\transparent[1]{}%
  }%
  \providecommand\rotatebox[2]{#2}%
  \ifx\svgwidth\undefined%
    \setlength{\unitlength}{230.0399905bp}%
    \ifx\svgscale\undefined%
      \relax%
    \else%
      \setlength{\unitlength}{\unitlength * \real{\svgscale}}%
    \fi%
  \else%
    \setlength{\unitlength}{\svgwidth}%
  \fi%
  \global\let\svgwidth\undefined%
  \global\let\svgscale\undefined%
  \makeatother%
  \begin{picture}(1,0.39530342)%
    \put(0,0){\includegraphics[width=\unitlength,page=1]{skip_example_v2.pdf}}%
    \put(0.02077898,0.01243267){\color[rgb]{0,0,0}\makebox(0,0)[lb]{\smash{a}}}%
    \put(0,0){\includegraphics[width=\unitlength,page=2]{skip_example_v2.pdf}}%
    \put(0.25747695,0.01243267){\color[rgb]{0,0,0}\makebox(0,0)[lb]{\smash{b}}}%
    \put(0,0){\includegraphics[width=\unitlength,page=3]{skip_example_v2.pdf}}%
    \put(0.4903495,0.01243267){\color[rgb]{0,0,0}\makebox(0,0)[lb]{\smash{c}}}%
    \put(0,0){\includegraphics[width=\unitlength,page=4]{skip_example_v2.pdf}}%
    \put(0.72404799,0.01243267){\color[rgb]{0,0,0}\makebox(0,0)[lb]{\smash{d}}}%
    \put(0,0){\includegraphics[width=\unitlength,page=5]{skip_example_v2.pdf}}%
    \put(0.95992002,0.01243267){\color[rgb]{0,0,0}\makebox(0,0)[lb]{\smash{e}}}%
    \put(0,0){\includegraphics[width=\unitlength,page=6]{skip_example_v2.pdf}}%
    \put(0.04121021,0.36524088){\color[rgb]{0,0,0}\makebox(0,0)[lb]{\smash{Contiguous}}}%
    \put(0.04121021,0.31881417){\color[rgb]{0,0,0}\makebox(0,0)[lb]{\smash{Skip}}}%
    \put(0,0){\includegraphics[width=\unitlength,page=10]{skip_example_v2.pdf}}%
    \put(0.12890417,0.14062777){\color[rgb]{0,0,0}\makebox(0,0)[lb]{\smash{ab}}}%
    \put(0.24463108,0.20583374){\color[rgb]{0,0,0}\makebox(0,0)[lb]{\smash{ac}}}%
    \put(0.36002566,0.27104003){\color[rgb]{0,0,0}\makebox(0,0)[lb]{\smash{ad}}}%
    \put(0.47927803,0.34276665){\color[rgb]{0,0,0}\makebox(0,0)[lb]{\smash{ae}}}%
    \put(0,0){\includegraphics[width=\unitlength,page=8]{skip_example_v2.pdf}}%
    \put(0,0){\includegraphics[width=\unitlength,page=12]{skip_example_v2.pdf}}%
  \end{picture}%
\endgroup%

%% file: pipeline.pdf_tex
\begingroup%
  \makeatletter%
  \providecommand\color[2][]{%
    \errmessage{(Inkscape) Color is used for the text in Inkscape, but the package 'color.sty' is not loaded}%
    \renewcommand\color[2][]{}%
  }%
  \providecommand\transparent[1]{%
    \errmessage{(Inkscape) Transparency is used (non-zero) for the text in Inkscape, but the package 'transparent.sty' is not loaded}%
    \renewcommand\transparent[1]{}%
  }%
  \providecommand\rotatebox[2]{#2}%
  \ifx\svgwidth\undefined%
    \setlength{\unitlength}{384.59998108bp}%
    \ifx\svgscale\undefined%
      \relax%
    \else%
      \setlength{\unitlength}{\unitlength * \real{\svgscale}}%
    \fi%
  \else%
    \setlength{\unitlength}{\svgwidth}%
  \fi%
  \global\let\svgwidth\undefined%
  \global\let\svgscale\undefined%
  \makeatother%
  \begin{picture}(1,0.85759503)%
    \put(0,0){\includegraphics[width=\unitlength,page=1]{pipeline.pdf}}%
    \put(0.02802806,0.76861678){\color[rgb]{0,0,0}\makebox(0,0)[lb]{\smash{Fixed (\#)}}}%
    \put(0,0){\includegraphics[width=\unitlength,page=2]{pipeline.pdf}}%
    \put(0.01791885,0.23333335){\color[rgb]{0,0,0}\makebox(0,0)[lb]{\smash{Variable (s)}}}%
    \put(0,0){\includegraphics[width=\unitlength,page=3]{pipeline.pdf}}%
    \put(0.35208787,0.57535104){\color[rgb]{0,0,0}\makebox(0,0)[lb]{\smash{Count*}}}%
    \put(0,0){\includegraphics[width=\unitlength,page=4]{pipeline.pdf}}%
    \put(0.32724283,0.48611547){\color[rgb]{0,0,0}\makebox(0,0)[lb]{\smash{Periodicity}}}%
    \put(0,0){\includegraphics[width=\unitlength,page=5]{pipeline.pdf}}%
    \put(0.32960893,0.3968799){\color[rgb]{0,0,0}\makebox(0,0)[lb]{\smash{Resonance}}}%
    \put(0,0){\includegraphics[width=\unitlength,page=6]{pipeline.pdf}}%
    \put(0.33116899,0.30769633){\color[rgb]{0,0,0}\makebox(0,0)[lb]{\smash{Proximity}}}%
    \put(0,0){\includegraphics[width=\unitlength,page=7]{pipeline.pdf}}%
    \put(0.33616119,0.23406138){\color[rgb]{0,0,0}\makebox(0,0)[lb]{\smash{Resonant}}}%
    \put(0.32742484,0.20286014){\color[rgb]{0,0,0}\makebox(0,0)[lb]{\smash{Periodicity}}}%
    \put(0.62242069,0.53073326){\color[rgb]{0,0,0}\makebox(0,0)[lb]{\smash{None*}}}%
    \put(0,0){\includegraphics[width=\unitlength,page=9]{pipeline.pdf}}%
    \put(0.59781954,0.44149769){\color[rgb]{0,0,0}\makebox(0,0)[lb]{\smash{Frequency}}}%
    \put(0,0){\includegraphics[width=\unitlength,page=10]{pipeline.pdf}}%
    \put(0.60043577,0.35226212){\color[rgb]{0,0,0}\makebox(0,0)[lb]{\smash{Harmony}}}%
    \put(0,0){\includegraphics[width=\unitlength,page=11]{pipeline.pdf}}%
    \put(0.62239677,0.26307855){\color[rgb]{0,0,0}\makebox(0,0)[lb]{\smash{Both}}}%
    \put(0,0){\includegraphics[width=\unitlength,page=12]{pipeline.pdf}}%
    \put(0.89710972,0.66253253){\color[rgb]{0,0,0}\makebox(0,0)[lb]{\smash{Count*}}}%
    \put(0,0){\includegraphics[width=\unitlength,page=13]{pipeline.pdf}}%
    \put(0.90091003,0.57334896){\color[rgb]{0,0,0}\makebox(0,0)[lb]{\smash{$PMI$}}}%
    \put(0,0){\includegraphics[width=\unitlength,page=14]{pipeline.pdf}}%
    \put(0.88442537,0.48411339){\color[rgb]{0,0,0}\makebox(0,0)[lb]{\smash{$PMI_{\text{local}}$}}}%
    \put(0,0){\includegraphics[width=\unitlength,page=15]{pipeline.pdf}}%
    \put(0.15613102,0.70234013){\color[rgb]{0,0,0}\makebox(0,0)[lb]{\smash{0*}}}%
    \put(0.15613102,0.65317216){\color[rgb]{0,0,0}\makebox(0,0)[lb]{\smash{1}}}%
    \put(0,0){\includegraphics[width=\unitlength,page=17]{pipeline.pdf}}%
    \put(0.15613102,0.60405619){\color[rgb]{0,0,0}\makebox(0,0)[lb]{\smash{2}}}%
    \put(0,0){\includegraphics[width=\unitlength,page=18]{pipeline.pdf}}%
    \put(0.15613102,0.55488823){\color[rgb]{0,0,0}\makebox(0,0)[lb]{\smash{3}}}%
    \put(0,0){\includegraphics[width=\unitlength,page=19]{pipeline.pdf}}%
    \put(0.15613102,0.50577226){\color[rgb]{0,0,0}\makebox(0,0)[lb]{\smash{4}}}%
    \put(0,0){\includegraphics[width=\unitlength,page=20]{pipeline.pdf}}%
    \put(0.15613102,0.45660429){\color[rgb]{0,0,0}\makebox(0,0)[lb]{\smash{5}}}%
    \put(0,0){\includegraphics[width=\unitlength,page=21]{pipeline.pdf}}%
    \put(0.15613102,0.40748833){\color[rgb]{0,0,0}\makebox(0,0)[lb]{\smash{6}}}%
    \put(0,0){\includegraphics[width=\unitlength,page=22]{pipeline.pdf}}%
    \put(0.15613102,0.35832036){\color[rgb]{0,0,0}\makebox(0,0)[lb]{\smash{7}}}%
    \put(0,0){\includegraphics[width=\unitlength,page=23]{pipeline.pdf}}%
    \put(0.15613102,0.30915239){\color[rgb]{0,0,0}\makebox(0,0)[lb]{\smash{8}}}%
    \put(0,0){\includegraphics[width=\unitlength,page=24]{pipeline.pdf}}%
    \put(0.14664584,0.16050444){\color[rgb]{0,0,0}\makebox(0,0)[lb]{\smash{0.5}}}%
    \put(0,0){\includegraphics[width=\unitlength,page=25]{pipeline.pdf}}%
    \put(0.14664584,0.11089445){\color[rgb]{0,0,0}\makebox(0,0)[lb]{\smash{1.0}}}%
    \put(0,0){\includegraphics[width=\unitlength,page=26]{pipeline.pdf}}%
    \put(0.14664584,0.06133647){\color[rgb]{0,0,0}\makebox(0,0)[lb]{\smash{1.5}}}%
    \put(0,0){\includegraphics[width=\unitlength,page=27]{pipeline.pdf}}%
    \put(0.14664584,0.01180449){\color[rgb]{0,0,0}\makebox(0,0)[lb]{\smash{2.0}}}%
    \put(0,0){\includegraphics[width=\unitlength,page=28]{pipeline.pdf}}%
    \put(0.05965676,0.8340614){\color[rgb]{0,0,0}\makebox(0,0)[lb]{\smash{\textbf{Skip}}}}%
    \put(0.35108787,0.8340614){\color[rgb]{0,0,0}\makebox(0,0)[lb]{\smash{\textbf{Count}}}}%
    \put(0.61419604,0.8340614){\color[rgb]{0,0,0}\makebox(0,0)[lb]{\smash{\textbf{Filter}}}}%
    \put(0.89585543,0.8340614){\color[rgb]{0,0,0}\makebox(0,0)[lb]{\smash{\textbf{Rank}}}}%
    \put(0,0){\includegraphics[width=\unitlength,page=29]{pipeline.pdf}}%
    \put(0.86864482,0.3968799){\color[rgb]{0,0,0}\makebox(0,0)[lb]{\smash{$PMI_{\text{coverage}}$}}}%
    \put(0,0){\includegraphics[width=\unitlength,page=30]{pipeline.pdf}}%
    \put(0.90578211,0.30327615){\color[rgb]{0,0,0}\makebox(0,0)[lb]{\smash{$Dice$}}}%
    \put(0.91819865,0.20967241){\color[rgb]{0,0,0}\makebox(0,0)[lb]{\smash{$\chi^2$}}}%
    \put(0.91507852,0.12298494){\color[rgb]{0,0,0}\makebox(0,0)[lb]{\smash{$G^2$}}}%
    \put(0,0){\includegraphics[width=\unitlength,page=34]{pipeline.pdf}}%
    \put(0,0){\includegraphics[width=\unitlength,page=36]{pipeline.pdf}}%
    \put(0,0){\includegraphics[width=\unitlength,page=37]{pipeline.pdf}}%
    \put(0,0){\includegraphics[width=\unitlength,page=39]{pipeline.pdf}}%
    \put(0,0){\includegraphics[width=\unitlength,page=41]{pipeline.pdf}}%
  \end{picture}%
\endgroup%

%% file: RR_fig1_rev.pdf_tex
\begingroup%
  \makeatletter%
  \providecommand\color[2][]{%
    \errmessage{(Inkscape) Color is used for the text in Inkscape, but the package 'color.sty' is not loaded}%
    \renewcommand\color[2][]{}%
  }%
  \providecommand\transparent[1]{%
    \errmessage{(Inkscape) Transparency is used (non-zero) for the text in Inkscape, but the package 'transparent.sty' is not loaded}%
    \renewcommand\transparent[1]{}%
  }%
  \providecommand\rotatebox[2]{#2}%
  \ifx\svgwidth\undefined%
    \setlength{\unitlength}{418.36535181bp}%
    \ifx\svgscale\undefined%
      \relax%
    \else%
      \setlength{\unitlength}{\unitlength * \real{\svgscale}}%
    \fi%
  \else%
    \setlength{\unitlength}{\svgwidth}%
  \fi%
  \global\let\svgwidth\undefined%
  \global\let\svgscale\undefined%
  \makeatother%
  \begin{picture}(1,0.60345698)%
    \put(0,.004){\includegraphics[width=\unitlength,page=1]{RR_fig1_rev.pdf}}%
    \put(0.11059304,0.17067813){\makebox(0,0)[b]{\smash{Fixed}}}%
    \put(0.24231947,0.17067813){\makebox(0,0)[b]{\smash{Variable}}}%
    \put(0.17645643,0.15291147){\makebox(0,0)[b]{\smash{\textbf{Skip Type}}}}%
    \put(0,.004){\includegraphics[width=\unitlength,page=2]{RR_fig1_rev.pdf}}%
    \put(0.04009261,0.18088){\makebox(0,0)[rb]{\smash{0}}}%
    \put(0.04009261,0.28591336){\makebox(0,0)[rb]{\smash{.01}}}%
    \put(0.04009261,0.39094671){\makebox(0,0)[rb]{\smash{.02}}}%
    \put(0.0084435,0.31564979){\rotatebox{90}{\makebox(0,0)[b]{\smash{\textbf{\textit{MRR}}}}}}%
    \put(0,.004){\includegraphics[width=\unitlength,page=3]{RR_fig1_rev.pdf}}%
    \put(0.42459658,0.32370684){\makebox(0,0)[b]{\smash{None}}}%
    \put(0.49222292,0.32370684){\makebox(0,0)[b]{\smash{1}}}%
    \put(0.55984927,0.32370684){\makebox(0,0)[b]{\smash{2}}}%
    \put(0.62747554,0.32370684){\makebox(0,0)[b]{\smash{3}}}%
    \put(0.69510189,0.32370684){\makebox(0,0)[b]{\smash{4}}}%
    \put(0.76272823,0.32370684){\makebox(0,0)[b]{\smash{5}}}%
    \put(0.83035449,0.32370684){\makebox(0,0)[b]{\smash{6}}}%
    \put(0.89798093,0.32370684){\makebox(0,0)[b]{\smash{7}}}%
    \put(0.96560718,0.32370684){\makebox(0,0)[b]{\smash{8}}}%
    \put(0.69510214,0.30594019){\makebox(0,0)[b]{\smash{\textbf{Boundary (skips)}}}}%
    \put(0,.004){\includegraphics[width=\unitlength,page=4]{RR_fig1_rev.pdf}}%
    \put(0.38614621,0.33390873){\makebox(0,0)[rb]{\smash{0}}}%
    \put(0.38614621,0.3862515){\makebox(0,0)[rb]{\smash{.01}}}%
    \put(0.38614621,0.43859428){\makebox(0,0)[rb]{\smash{.02}}}%
    \put(0.38614621,0.49093705){\makebox(0,0)[rb]{\smash{.03}}}%
    \put(0.38614621,0.54327984){\makebox(0,0)[rb]{\smash{.04}}}%
    \put(0.38614621,0.59562261){\makebox(0,0)[rb]{\smash{.05}}}%
    \put(0.35449708,0.46824376){\rotatebox{90}{\makebox(0,0)[b]{\smash{\textbf{\textit{MRR}}}}}}%
    \put(0,.004){\includegraphics[width=\unitlength,page=5]{RR_fig1_rev.pdf}}%
    \put(0.45164711,0.01851888){\makebox(0,0)[b]{\smash{None}}}%
    \put(0.5733745,0.01851888){\makebox(0,0)[b]{\smash{0.5}}}%
    \put(0.69510189,0.01851888){\makebox(0,0)[b]{\smash{1}}}%
    \put(0.81682928,0.01851888){\makebox(0,0)[b]{\smash{1.5}}}%
    \put(0.93855667,0.01851888){\makebox(0,0)[b]{\smash{2}}}%
    \put(0.69510214,0.00075223){\makebox(0,0)[b]{\smash{\textbf{Boundary (s)}}}}%
    \put(0,.004){\includegraphics[width=\unitlength,page=6]{RR_fig1_rev.pdf}}%
    \put(0.38614621,0.02872077){\makebox(0,0)[rb]{\smash{0}}}%
    \put(0.38614621,0.08106355){\makebox(0,0)[rb]{\smash{.01}}}%
    \put(0.38614621,0.13340632){\makebox(0,0)[rb]{\smash{.02}}}%
    \put(0.38614621,0.18574909){\makebox(0,0)[rb]{\smash{.03}}}%
    \put(0.38614621,0.23809188){\makebox(0,0)[rb]{\smash{.04}}}%
    \put(0.38614621,0.29043466){\makebox(0,0)[rb]{\smash{.05}}}%
    \put(0.35449708,0.16305573){\rotatebox{90}{\makebox(0,0)[b]{\smash{\textbf{\textit{MRR}}}}}}%
    \put(0,.004){\includegraphics[width=\unitlength,page=7]{RR_fig1_rev.pdf}}%
    \put(0.41008844,0.58392411){\color[rgb]{0,0,0}\makebox(0,0)[lb]{\smash{\textit{Fixed}}}}%
    \put(0.41023752,0.27588793){\color[rgb]{0,0,0}\makebox(0,0)[lb]{\smash{\textit{Variable}}}}%
    \put(0,.004){\includegraphics[width=\unitlength,page=8]{RR_fig1_rev.pdf}}%
    \put(0.57560072,0.56960014){\color[rgb]{0,0,0}\makebox(0,0)[lb]{\smash{***}}}%
    \put(0,.004){\includegraphics[width=\unitlength,page=9]{RR_fig1_rev.pdf}}%
    \put(0.68318623,0.19020529){\color[rgb]{0,0,0}\makebox(0,0)[lb]{\smash{**}}}%
    \put(0,.004){\includegraphics[width=\unitlength,page=10]{RR_fig1_rev.pdf}}%
    \put(0.16908151,0.4292856){\color[rgb]{0,0,0}\makebox(0,0)[lb]{\smash{*}}}%
  \end{picture}%
\endgroup%

%% file: RR_fig2_rev.pdf_tex
\begingroup%
  \makeatletter%
  \providecommand\color[2][]{%
    \errmessage{(Inkscape) Color is used for the text in Inkscape, but the package 'color.sty' is not loaded}%
    \renewcommand\color[2][]{}%
  }%
  \providecommand\transparent[1]{%
    \errmessage{(Inkscape) Transparency is used (non-zero) for the text in Inkscape, but the package 'transparent.sty' is not loaded}%
    \renewcommand\transparent[1]{}%
  }%
  \providecommand\rotatebox[2]{#2}%
  \ifx\svgwidth\undefined%
    \setlength{\unitlength}{418.36535181bp}%
    \ifx\svgscale\undefined%
      \relax%
    \else%
      \setlength{\unitlength}{\unitlength * \real{\svgscale}}%
    \fi%
  \else%
    \setlength{\unitlength}{\svgwidth}%
  \fi%
  \global\let\svgwidth\undefined%
  \global\let\svgscale\undefined%
  \makeatother%
  \begin{picture}(1,0.61508591)%
    \put(0,.006){\includegraphics[width=\unitlength,page=1]{RR_fig2_rev.pdf}}%
    \put(0.34330972,0.34805231){\makebox(0,0)[b]{\smash{Count}}}%
    \put(0.43269241,0.34805231){\makebox(0,0)[b]{\smash{Per.}}}%
    \put(0.52207509,0.34805231){\makebox(0,0)[b]{\smash{Res.}}}%
    \put(0.61145777,0.34805231){\makebox(0,0)[b]{\smash{Prox.}}}%
    \put(0.70084045,0.34805231){\makebox(0,0)[b]{\smash{Per. \!$\times$\! Res.}}}%
    \put(0.52207527,0.32828566){\makebox(0,0)[b]{\smash{\textbf{Count Type}}}}%
    \put(0,.006){\includegraphics[width=\unitlength,page=2]{RR_fig2_rev.pdf}}%
    \put(0.29398118,0.3582542){\makebox(0,0)[rb]{\smash{0}}}%
    \put(0.29398118,0.40572788){\makebox(0,0)[rb]{\smash{.005}}}%
    \put(0.29398118,0.45320156){\makebox(0,0)[rb]{\smash{.01}}}%
    \put(0.29398118,0.50067525){\makebox(0,0)[rb]{\smash{.015}}}%
    \put(0.29398118,0.54814893){\makebox(0,0)[rb]{\smash{.02}}}%
    \put(0.29398118,0.59562261){\makebox(0,0)[rb]{\smash{.025}}}%
    \put(0.25633207,0.48041642){\rotatebox{90}{\makebox(0,0)[b]{\smash{\textbf{\textit{MRR}}}}}}%
    \put(0,.006){\includegraphics[width=\unitlength,page=3]{RR_fig2_rev.pdf}}%
    \put(0.100594,0.01851888){\makebox(0,0)[b]{\smash{None}}}%
    \put(0.21232236,0.01851888){\makebox(0,0)[b]{\smash{Freq.}}}%
    \put(0.32405071,0.01851888){\makebox(0,0)[b]{\smash{Harm.}}}%
    \put(0.43577906,0.01851888){\makebox(0,0)[b]{\smash{Both}}}%
    \put(0.26818672,-0.00275223){\makebox(0,0)[b]{\smash{\textbf{Filter Type}}}}%
    \put(0,.006){\includegraphics[width=\unitlength,page=4]{RR_fig2_rev.pdf}}%
    \put(0.04009261,0.02872077){\makebox(0,0)[rb]{\smash{0}}}%
    \put(0.04009261,0.05509501){\makebox(0,0)[rb]{\smash{.005}}}%
    \put(0.04009261,0.08146934){\makebox(0,0)[rb]{\smash{.01}}}%
    \put(0.04009261,0.10784358){\makebox(0,0)[rb]{\smash{.015}}}%
    \put(0.04009261,0.13421782){\makebox(0,0)[rb]{\smash{.02}}}%
    \put(0.04009261,0.16059213){\makebox(0,0)[rb]{\smash{.025}}}%
    \put(0.04009261,0.18696638){\makebox(0,0)[rb]{\smash{.03}}}%
    \put(0.04009261,0.21334062){\makebox(0,0)[rb]{\smash{.035}}}%
    \put(0.04009261,0.23971493){\makebox(0,0)[rb]{\smash{.04}}}%
    \put(0.04009261,0.26608917){\makebox(0,0)[rb]{\smash{.045}}}%
    \put(0.0004435,0.15088298){\rotatebox{90}{\makebox(0,0)[b]{\smash{\textbf{\textit{MRR}}}}}}%
    \put(0,.006){\includegraphics[width=\unitlength,page=5]{RR_fig2_rev.pdf}}%
    \put(0.58442934,0.01851888){\makebox(0,0)[b]{\smash{Count}}}%
    \put(0.64827413,0.01851888){\makebox(0,0)[b]{\smash{PMI}}}%
    \put(0.71211885,0.01851888){\makebox(0,0)[b]{\smash{PMI$_{\text{loc}}$}}}%
    \put(0.77596366,0.01851888){\makebox(0,0)[b]{\smash{PMI$_{\text{cov}}$}}}%
    \put(0.83980845,0.01851888){\makebox(0,0)[b]{\smash{Dice}}}%
    \put(0.90365325,0.01851888){\makebox(0,0)[b]{\smash{$\chi^2$}}}%
    \put(0.96749804,0.01851888){\makebox(0,0)[b]{\smash{$G^2$}}}%
    \put(0.77596382,-0.00275223){\makebox(0,0)[b]{\smash{\textbf{AM Type}}}}%
    \put(0,.006){\includegraphics[width=\unitlength,page=6]{RR_fig2_rev.pdf}}%
    \put(0.54786973,0.02872077){\makebox(0,0)[rb]{\smash{0}}}%
    \put(0.54786973,0.07619445){\makebox(0,0)[rb]{\smash{.02}}}%
    \put(0.54786973,0.12366813){\makebox(0,0)[rb]{\smash{.04}}}%
    \put(0.54786973,0.17114181){\makebox(0,0)[rb]{\smash{.06}}}%
    \put(0.54786973,0.21861549){\makebox(0,0)[rb]{\smash{.08}}}%
    \put(0.54783924,0.26608917){\makebox(0,0)[rb]{\smash{.10}}}%
    \put(0.51622062,0.15088298){\rotatebox{90}{\makebox(0,0)[b]{\smash{\textbf{\textit{MRR}}}}}}%
    \put(0,.006){\includegraphics[width=\unitlength,page=7]{RR_fig2_rev.pdf}}%
    \put(0.37122749,0.61108){\color[rgb]{0,0,0}\makebox(0,0)[lb]{\smash{n.s.}}}%
    \put(0,.006){\includegraphics[width=\unitlength,page=8]{RR_fig2_rev.pdf}}%
    \put(0.19318114,0.26740095){\color[rgb]{0,0,0}\makebox(0,0)[lb]{\smash{***}}}%
    \put(0,.006){\includegraphics[width=\unitlength,page=9]{RR_fig2_rev.pdf}}%
    \put(0.66101803,0.27970896){\color[rgb]{0,0,0}\makebox(0,0)[lb]{\smash{***}}}%
  \end{picture}%
\endgroup%

%% file: top10_v2.pdf_tex
\begingroup%
  \makeatletter%
  \providecommand\color[2][]{%
    \errmessage{(Inkscape) Color is used for the text in Inkscape, but the package 'color.sty' is not loaded}%
    \renewcommand\color[2][]{}%
  }%
  \providecommand\transparent[1]{%
    \errmessage{(Inkscape) Transparency is used (non-zero) for the text in Inkscape, but the package 'transparent.sty' is not loaded}%
    \renewcommand\transparent[1]{}%
  }%
  \providecommand\rotatebox[2]{#2}%
  \ifx\svgwidth\undefined%
    \setlength{\unitlength}{180.20521912bp}%
    \ifx\svgscale\undefined%
      \relax%
    \else%
      \setlength{\unitlength}{\unitlength * \real{\svgscale}}%
    \fi%
  \else%
    \setlength{\unitlength}{\svgwidth}%
  \fi%
  \global\let\svgwidth\undefined%
  \global\let\svgscale\undefined%
  \makeatother%
  \begin{picture}(1,1.89573607)%
    \put(0.11471456,1.96715928){\color[rgb]{0,0,0}\makebox(0,0)[lt]{\begin{minipage}{0.08733093\unitlength}\raggedright \end{minipage}}}%
    \put(-0.00224877,1.46346952){\color[rgb]{0,0,0}\makebox(0,0)[lb]{\smash{3}}}
    \put(0.10244095,1.16375138){\color[rgb]{0,0,0}\makebox(0,0)[lb]{\smash{V($^6_4$)}}}%
    \put(0.35070259,1.16291247){\color[rgb]{0,0,0}\makebox(0,0)[lb]{\smash{I}}}%
    \put(0.22149686,1.16375189){\color[rgb]{0,0,0}\makebox(0,0)[lb]{\smash{V$^7$}}}%
    \put(0,0){\includegraphics[width=\unitlength,page=1]{top10_v2.pdf}}%
    \put(0.50176005,1.46398451){\color[rgb]{0,0,0}\makebox(0,0)[lb]{\smash{4}}}
    \put(-0.0026552,1.08524681){\color[rgb]{0,0,0}\makebox(0,0)[lb]{\smash{5}}}
    \put(0.10380306,0.78325761){\color[rgb]{0,0,0}\makebox(0,0)[lb]{\smash{V}}}%
    \put(0.34250645,0.78241866){\color[rgb]{0,0,0}\makebox(0,0)[lb]{\smash{V}}}%
    \put(0.23384408,0.78325812){\color[rgb]{0,0,0}\makebox(0,0)[lb]{\smash{I}}}%
    \put(0.50024269,1.08552556){\color[rgb]{0,0,0}\makebox(0,0)[lb]{\smash{6}}}
    \put(0.60477198,1.54318397){\color[rgb]{0,0,0}\makebox(0,0)[lb]{\smash{V($^6_4$)}}}%
    \put(0.85303367,1.54234497){\color[rgb]{0,0,0}\makebox(0,0)[lb]{\smash{I}}}%
    \put(0.7238279,1.54318444){\color[rgb]{0,0,0}\makebox(0,0)[lb]{\smash{V$^7$}}}%
    \put(0,0){\includegraphics[width=\unitlength,page=2]{top10_v2.pdf}}%
    \put(0.49926724,0.69517326){\color[rgb]{0,0,0}\makebox(0,0)[lb]{\smash{8}}}
    \put(0.60823093,0.39594412){\color[rgb]{0,0,0}\makebox(0,0)[lb]{\smash{ii$^6$}}}%
    \put(0.72344966,0.39662592){\color[rgb]{0,0,0}\makebox(0,0)[lb]{\smash{V($^6_4$)}}}%
    \put(0.84250558,0.39662659){\color[rgb]{0,0,0}\makebox(0,0)[lb]{\smash{V$^7$}}}%
    \put(-0.00219456,0.31628996){\color[rgb]{0,0,0}\makebox(0,0)[lb]{\smash{9}}}
    \put(0.11131341,0.01209203){\color[rgb]{0,0,0}\makebox(0,0)[lb]{\smash{I$^6$}}}%
    \put(0.22091847,0.01206442){\color[rgb]{0,0,0}\makebox(0,0)[lb]{\smash{vi$^6$}}}%
    \put(0.33997439,0.0120651){\color[rgb]{0,0,0}\makebox(0,0)[lb]{\smash{V$^6$}}}%
    \put(0.49677167,0.31560985){\color[rgb]{0,0,0}\makebox(0,0)[lb]{\smash{10}}}
    \put(-0.00205908,0.70079031){\color[rgb]{0,0,0}\makebox(0,0)[lb]{\smash{7}}}
    \put(-0.00650281,1.8467737){\color[rgb]{0,0,0}\makebox(0,0)[lb]{\smash{1}}}
    \put(0,0){\includegraphics[width=\unitlength,page=3]{top10_v2.pdf}}%
    \put(0.10244188,1.54356911){\color[rgb]{0,0,0}\makebox(0,0)[lb]{\smash{V$^7$}}}%
    \put(0.35070354,1.54273012){\color[rgb]{0,0,0}\makebox(0,0)[lb]{\smash{I}}}%
    \put(0,0){\includegraphics[width=\unitlength,page=4]{top10_v2.pdf}}%
    \put(0.50143207,1.84683165){\color[rgb]{0,0,0}\makebox(0,0)[lb]{\smash{2}}}
    \put(0.60995346,1.16091144){\color[rgb]{0,0,0}\makebox(0,0)[lb]{\smash{V$^6$}}}%
    \put(0.85821515,1.1600724){\color[rgb]{0,0,0}\makebox(0,0)[lb]{\smash{I}}}%
    \put(0.72284379,1.16091144){\color[rgb]{0,0,0}\makebox(0,0)[lb]{\smash{$^5$}}}%
    \put(0,0){\includegraphics[width=\unitlength,page=5]{top10_v2.pdf}}%
    \put(0.60823034,0.78582088){\color[rgb]{0,0,0}\makebox(0,0)[lb]{\smash{ii$^6$}}}%
    \put(0.73573999,0.78577685){\color[rgb]{0,0,0}\makebox(0,0)[lb]{\smash{I$^6$}}}%
    \put(0.83501341,0.78577685){\color[rgb]{0,0,0}\makebox(0,0)[lb]{\smash{vii$^6$}}}%
    \put(0.09762362,0.39591109){\color[rgb]{0,0,0}\makebox(0,0)[lb]{\smash{iii$^6$}}}%
    \put(0.22593524,0.39586706){\color[rgb]{0,0,0}\makebox(0,0)[lb]{\smash{ii$^6$}}}%
    \put(0.35327748,0.39586706){\color[rgb]{0,0,0}\makebox(0,0)[lb]{\smash{I$^6$}}}%
    \put(0.60216726,0.01197636){\color[rgb]{0,0,0}\makebox(0,0)[lb]{\smash{V($^6_4$)}}}%
    \put(0.85439837,0.01057026){\color[rgb]{0,0,0}\makebox(0,0)[lb]{\smash{I}}}%
    \put(0.72122314,0.01197687){\color[rgb]{0,0,0}\makebox(0,0)[lb]{\smash{V}}}%
    \put(0,0){\includegraphics[width=\unitlength,page=6]{top10_v2.pdf}}%
    \put(0.1164868,1.85273125){\color[rgb]{0,0,0}\makebox(0,0)[lb]{\smash{\tiny $\hat{3}$}}}%
    \put(0,0){\includegraphics[width=\unitlength,page=7]{top10_v2.pdf}}%
    \put(0.2348472,1.84370455){\color[rgb]{0,0,0}\makebox(0,0)[lb]{\smash{\tiny$\hat{2}$}}}%
    \put(0,0){\includegraphics[width=\unitlength,page=8]{top10_v2.pdf}}%
    \put(0.36121903,1.83746354){\color[rgb]{0,0,0}\makebox(0,0)[lb]{\smash{\tiny $\hat{1}$}}}%
    \put(0,0){\includegraphics[width=\unitlength,page=9]{top10_v2.pdf}}%
    \put(0.61678264,1.83741251){\color[rgb]{0,0,0}\makebox(0,0)[lb]{\smash{\tiny $\hat{1}$}}}%
    \put(0,0){\includegraphics[width=\unitlength,page=10]{top10_v2.pdf}}%
    \put(0.73648714,1.84374276){\color[rgb]{0,0,0}\makebox(0,0)[lb]{\smash{\tiny $\hat{2}$}}}%
    \put(0,0){\includegraphics[width=\unitlength,page=11]{top10_v2.pdf}}%
    \put(0.85548603,1.83741251){\color[rgb]{0,0,0}\makebox(0,0)[lb]{\smash{\tiny$\hat{1}$}}}%
    \put(0,0){\includegraphics[width=\unitlength,page=12]{top10_v2.pdf}}%
    \put(0.85761719,1.46960137){\color[rgb]{0,0,0}\makebox(0,0)[lb]{\smash{\tiny$\hat{3}$}}}%
    \put(0,0){\includegraphics[width=\unitlength,page=13]{top10_v2.pdf}}%
    \put(0.73818757,1.47114655){\color[rgb]{0,0,0}\makebox(0,0)[lb]{\smash{\tiny $\hat{4}$}}}%
    \put(0,0){\includegraphics[width=\unitlength,page=14]{top10_v2.pdf}}%
    \put(0.61928091,1.47741638){\color[rgb]{0,0,0}\makebox(0,0)[lb]{\smash{\tiny $\hat{5}$}}}%
    \put(0,0){\includegraphics[width=\unitlength,page=15]{top10_v2.pdf}}%
    \put(0.1150821,1.46957406){\color[rgb]{0,0,0}\makebox(0,0)[lb]{\smash{\tiny $\hat{3}$}}}%
    \put(0,0){\includegraphics[width=\unitlength,page=16]{top10_v2.pdf}}%
    \put(0.23391032,1.46223515){\color[rgb]{0,0,0}\makebox(0,0)[lb]{\smash{\tiny $\hat{2}$}}}%
    \put(0,0){\includegraphics[width=\unitlength,page=17]{top10_v2.pdf}}%
    \put(0.35290922,1.45467016){\color[rgb]{0,0,0}\makebox(0,0)[lb]{\smash{\tiny $\hat{1}$}}}%
    \put(0,0){\includegraphics[width=\unitlength,page=18]{top10_v2.pdf}}%
    \put(0.11596329,1.0824905){\color[rgb]{0,0,0}\makebox(0,0)[lb]{\smash{\tiny $\hat{2}$}}}%
    \put(0,0){\includegraphics[width=\unitlength,page=19]{top10_v2.pdf}}%
    \put(0.23493822,1.08938864){\color[rgb]{0,0,0}\makebox(0,0)[lb]{\smash{\tiny $\hat{3}$}}}%
    \put(0,0){\includegraphics[width=\unitlength,page=20]{top10_v2.pdf}}%
    \put(0.35431095,1.08252513){\color[rgb]{0,0,0}\makebox(0,0)[lb]{\smash{\tiny $\hat{2}$}}}%
    \put(0,0){\includegraphics[width=\unitlength,page=21]{top10_v2.pdf}}%
    \put(0.6185024,1.08252513){\color[rgb]{0,0,0}\makebox(0,0)[lb]{\smash{\tiny $\hat{2}$}}}%
    \put(0,0){\includegraphics[width=\unitlength,page=22]{top10_v2.pdf}}%
    \put(0.73792233,1.07496683){\color[rgb]{0,0,0}\makebox(0,0)[lb]{\smash{\tiny $\hat{1}$}}}%
    \put(0,0){\includegraphics[width=\unitlength,page=23]{top10_v2.pdf}}%
    \put(0.85689847,1.07279437){\color[rgb]{0,0,0}\makebox(0,0)[lb]{\smash{\tiny $\hat{7}$}}}%
    \put(0,0){\includegraphics[width=\unitlength,page=24]{top10_v2.pdf}}%
    \put(0.61787311,0.69700919){\color[rgb]{0,0,0}\makebox(0,0)[lb]{\smash{\tiny $\hat{2}$}}}%
    \put(0,0){\includegraphics[width=\unitlength,page=25]{top10_v2.pdf}}%
    \put(0.7372931,0.69022591){\color[rgb]{0,0,0}\makebox(0,0)[lb]{\smash{\tiny $\hat{1}$}}}%
    \put(0,0){\includegraphics[width=\unitlength,page=26]{top10_v2.pdf}}%
    \put(0.85657651,0.69690944){\color[rgb]{0,0,0}\makebox(0,0)[lb]{\smash{\tiny $\hat{2}$}}}%
    \put(0,0){\includegraphics[width=\unitlength,page=27]{top10_v2.pdf}}%
    \put(0.11605438,0.70579439){\color[rgb]{0,0,0}\makebox(0,0)[lb]{\smash{\tiny $\hat{3}$}}}%
    \put(0,0){\includegraphics[width=\unitlength,page=28]{top10_v2.pdf}}%
    \put(0.23542763,0.69698048){\color[rgb]{0,0,0}\makebox(0,0)[lb]{\smash{\tiny $\hat{2}$}}}%
    \put(0,0){\includegraphics[width=\unitlength,page=29]{top10_v2.pdf}}%
    \put(0.35437929,0.69021651){\color[rgb]{0,0,0}\makebox(0,0)[lb]{\smash{\tiny $\hat{1}$}}}%
    \put(0,0){\includegraphics[width=\unitlength,page=30]{top10_v2.pdf}}%
    \put(0.11536478,0.30765436){\color[rgb]{0,0,0}\makebox(0,0)[lb]{\smash{\tiny $\hat{1}$}}}%
    \put(0,0){\includegraphics[width=\unitlength,page=31]{top10_v2.pdf}}%
    \put(0.2335557,0.30100283){\color[rgb]{0,0,0}\makebox(0,0)[lb]{\smash{\tiny $\hat{6}$}}}%
    \put(0,0){\includegraphics[width=\unitlength,page=32]{top10_v2.pdf}}%
    \put(0.35368125,0.30133111){\color[rgb]{0,0,0}\makebox(0,0)[lb]{\smash{\tiny $\hat{5}$}}}%
    \put(0,0){\includegraphics[width=\unitlength,page=33]{top10_v2.pdf}}%
    \put(0.61738382,0.32143564){\color[rgb]{0,0,0}\makebox(0,0)[lb]{\smash{\tiny $\hat{3}$}}}%
    \put(0,0){\includegraphics[width=\unitlength,page=34]{top10_v2.pdf}}%
    \put(0.73722482,0.31341509){\color[rgb]{0,0,0}\makebox(0,0)[lb]{\smash{\tiny $\hat{2}$}}}%
    \put(0,0){\includegraphics[width=\unitlength,page=35]{top10_v2.pdf}}%
    \put(0.85664479,0.30766046){\color[rgb]{0,0,0}\makebox(0,0)[lb]{\smash{\tiny $\hat{1}$}}}%
    \put(0,0){\includegraphics[width=\unitlength,page=36]{top10_v2.pdf}}%
  \end{picture}%
\endgroup%